\documentclass[11pt,a4paper]{article}
\pdfoutput=1
\usepackage{jheppub}

\usepackage{color}
\usepackage{amsmath}
\usepackage{pifont}

\usepackage{bbm}
\usepackage{verbatim}   
\usepackage{subfigure}  
\usepackage{acronym}

\usepackage{amsfonts}
\usepackage{amssymb}
\usepackage{mathrsfs}
\usepackage{graphicx}
\usepackage{multirow}
 \usepackage{slashed}

\usepackage{algpseudocode}

\definecolor{mygreen}{RGB}{29,145,47}
\definecolor{mypurple}{RGB}{164,64,214}
\definecolor{myorange}{RGB}{199,146,32}

\newcommand{\veck}{\mathbf{k}}
\newcommand{\vecp}{\mathbf{p}}
\newcommand{\vecq}{\mathbf{q}}
\newcommand{\vecx}{\mathbf{x}}
\newcommand{\vecr}{\mathbf{r}}

\newcommand{\vecv}{\mathbf{v}}
\newcommand{\vecz}{\mathbf{z}}

\newcommand{\vecf}{\mathbf{f}}

\newcommand{\sun}{\odot}

\newcommand{\second}{{\, {\rm sec}}}
\newcommand{\fermi}{{\, {\rm fm}}}
\newcommand{\barn}{{\, {\rm barn}}}
\newcommand{\pb}{{\, {\rm pb}}}
\newcommand{\dy}{{\, {\rm day}}}
\newcommand{\yr}{{\, {\rm yr}}}
\newcommand{\Gyr}{{\, {\rm Gyr}}}

\newcommand{\km}{{\, {\rm km}}}

\newcommand{\cm}{{\, {\rm cm}}}

\newcommand{\keV}{{\, {\rm keV}}}
\newcommand{\MeV}{{\, {\rm MeV}}}
\newcommand{\GeV}{{\, {\rm GeV}}}
\newcommand{\TeV}{{\, {\rm TeV}}}
\newcommand{\kelvin}{{\, {\rm K}}}
\newcommand{\gram}{{\, {\rm g}}}

\def\beq{\begin{equation}}
\def\eeq{\end{equation}}
\def\bea{\begin{eqnarray}}
\def\eea{\end{eqnarray}}
\def\bitem{\begin{itemize}}
\def\eitem{\end{itemize}}
\newcommand{\bec}{\begin{center}}
\newcommand{\eec}{\end{center}}
\newcommand{\ba}{\begin{array}}
\newcommand{\ea}{\end{array}}

\def\bar#1{\overline{#1}}

\def\inv{^{\raise.15ex\hbox{${\scriptscriptstyle -}$}\kern-.05em 1}}
\def\lbar{{\lower.35ex\hbox{$\mathchar'26$}\mkern-10mu\lambda}} 

\let\<=\langle
\let\>=\rangle

\let\+=\uparrow

\begin{document}

\title{
Signatures of Large Composite Dark Matter States
} 
\author[a,b]{Edward Hardy,}
\emailAdd{ehardy@ictp.it}
\author[a]{Robert Lasenby,}
\emailAdd{robert.lasenby@physics.ox.ac.uk}
\author[a,c]{John March-Russell,}
\emailAdd{jmr@thphys.ox.ac.uk}
\author[d]{Stephen M. West}
\emailAdd{stephen.west@rhul.ac.uk}
\affiliation[a]{Rudolf Peierls Centre for Theoretical Physics, University of Oxford, 1 Keble Road,\\
Oxford, OX1 3NP, UK}
\affiliation[b]{Abdus Salam International Centre for Theoretical Physics,
Strada Costiera 11, 34151, Trieste, Italy}
\affiliation[c]{Stanford Institute for Theoretical Physics, Department of Physics, Stanford University,\\
Stanford, CA 94305, USA}
\affiliation[d]{Physics Department, Royal Holloway, University of London,
Egham, Surrey, TW20 0EX, UK}

\abstract{
We investigate the interactions of large composite dark matter (DM)
states with the Standard Model (SM) sector.  Elastic scattering with
SM nuclei can be coherently enhanced by factors as large as $A^2$,
where $A$ is the number of constituents in the composite state (there exist
models in which DM states of very large $A \gtrsim 10^8$ may be realised).
This enhancement, for a given direct detection event rate, weakens the expected
signals at colliders by up to $1/A$.   Moreover, the spatially extended
nature of the DM states leads to an additional, characteristic, form factor
modifying the momentum dependence of scattering processes, altering the recoil
energy spectra in direct detection experiments.  In particular, energy
recoil spectra with peaks and troughs are possible, and such features
could be confirmed with only $\mathcal{O}(50)$ events, independently
of the assumed halo velocity distribution. Large composite states also
generically give rise to low-energy collective excitations potentially relevant to
direct detection and indirect detection phenomenology.  We compute the form
factor for a generic class of such excitations---quantised surface modes---finding that they
can lead to coherently-enhanced, but generally sub-dominant, inelastic
scattering in direct detection experiments.  Finally, we study the modifications to capture rates
in astrophysical objects that follow from the elastic form factor, as well as the effects of
inelastic interactions between DM states once captured.   We argue that inelastic interactions
may lead to the DM collapsing to a dense configuration at the centre of the
object.
}

\maketitle


\section{Introduction}
\label{sec:intro}

Most models of dark matter (DM) assume that it can, for practical purposes,
be treated as a collection of point-like particles.  However, this is not necessarily the
case, and a variety of models in which DM is a composite state have been
proposed (for example, WIMPonium \cite{Pospelov2008,MarchRussell2008,Shepherd2009},
and dark atoms \cite{Kaplan:2009de}).  Given the centrality of DM to our present thoughts about
beyond-the-Standard-Model physics, and our present lack of knowledge concerning
many of its fundamental characteristics, and thus how DM may reveal itself in experiments
or observations, it is important to consider possible variations away from the standard
picture.   This is especially the case if there are qualitatively new features compared to traditional
DM models that may affect direct and/or indirect detection phenomenology.

In this paper,
we consider the consequences for present-day scattering processes
of a simple kind of compositeness, in which the DM states are composed
of a large number, $A$, of constituents, forming an extended semi-uniform object,
analogous to Standard Model (SM) nuclei formed out of constituent nucleons.\footnote{
The generic scattering phenomenology of small-number composite DM states has
been considered in~\cite{Laha:2013gva}.}
Models that realise such a scenario include Q-balls (non-topological solitons
carrying a conserved charge)
\cite{Frieman1988, Kusenko1997}, and Nuclear Dark Matter models, in which DM is made
up of bound states of strongly-interacting constituents with short-range
interactions \cite{Krnjaic2014,Detmold2014,ndm1}. The early-universe cosmology of these models
has interesting features, as investigated in \cite{Krnjaic2014,ndm1,Detmold2014,Kusenko1997}; here, we simply
assume that a late-time population of such states exists.

This large and extended compositeness affects elastic scattering and the associated
direct detection phenomenology, as we discuss in Section~\ref{sec:elastic}.
Large composite states containing $A$ constituents can have
low-momentum-transfer elastic scattering cross sections coherently
enhanced by a factor as large as $A^2$ \cite{Gelmini2002,Krnjaic2014,ndm1}. 
This case is realised if the size of the 
state is not too large compared to the inverse momentum exchange relevant in 
direct detection scatterings. In this situation, and
assuming that the mass of the composite state is $\propto A$,
so the DM number density is $\propto 1/A$, the event rate at direct detection
experiments will effectively be enhanced $\propto A$ for a given interaction
strength between SM and DM constituents.
Previous studies of the build-up of composite DM states in the early universe
show that large values of $A$ ($\gtrsim 10^8)$ could plausibly be realised \cite{ndm1}, so the effective
enhancement can be significant.

In addition, if a high enough proportion of the DM states have radii
larger than SM nuclei, but not so large as to significantly suppress
coherent scattering, the spatial extension of the DM states leads to a dark form
factor modifying the momentum dependence of DM-SM scattering
(as previously considered in~\cite{Gelmini2002}). In the
simplest case of a scalar interaction depending only on the density, this
form factor has a characteristic series of peaks and troughs, analogous to SM nuclear
form factors. In direct detection experiments with sufficiently good
energy resolution, these could lead to the striking signature of rises
in the energy recoil spectrum. We find that significant features of this kind could
be distinguished from point-like elastic scattering after the
observation of only $\mathcal{O}(50)$ events,
independently of assumptions about the DM halo velocity distribution.
A distribution over DM sizes may average out these peaks into a
smoothly-falling effective form factor, in many cases resembling that from
e.g.\ the exchange of an intermediate-mass mediator, but multiple direct
detection experiments using different SM nuclear targets would still
be able to separate out the momentum dependence from halo velocity
distribution effects, as with other models of dark form
factors \cite{Gelmini2002,Chang2009,Feldstein2009,McDermott:2011hx,Cherry2014}.

As we discuss in Section~\ref{sec:inelastic}, large composite states also generically give rise to long-wavelength
and low-energy collective excitations. There will also be higher-energy excitations,
and fission processes, but making the binding energy scale
low enough for these to be relevant in direct detection
experiments generally complicates the cosmology of such models~\cite{ndm1}. 
We focus on the most model-independent possibility of collective density excitations.
We find that they can be of sufficiently
low energy to be excited in collisions with SM nuclei, leading to
coherently-enhanced inelastic scattering.
Although this is sub-dominant to elastic scattering in the scenarios
we consider, a non-negligible fraction of events may be inelastic,
and we calculate the associated form factor in
leading approximation.\footnote{Taking the possibility of inelastic DM {\it self}-interactions further, we may also expect, in the context of galactic halos, exothermic DM-DM interactions, e.g, from fusions.  This can lead to an increase in the average kinetic energy of DM in the galactic halo, which might in turn result in a clearing out of the central high-density region.  This possibility was outlined in~\cite{ndm1}, and requires a dedicated study of halo dynamics.
In this work, we focus on the interactions of composite DM 
states with the SM, and leave the topic of modified halo dynamics to a future paper.}

As is well known, couplings to SM nuclei may also lead to the capture of DM by
astrophysical objects.  As discussed in Section~\ref{sec:capture}
we find that the composite nature of the DM changes 
the overall capture rate.  In addition, perhaps the most interesting qualitative
effect is that the large local DM density inside the star, and
the natural possibility of inelastic DM self-interactions, could lead to a DM distribution that 
is significantly modified compared to the naive expectation.  In particular,
for either dissipative collisions or fusions, there is the possibility
that this process runs away to a state in which most of the captured
DM lies in a single very dense configuration.\footnote{Since the large composite states with saturated densities
we consider in this paper require a short-range repulsive interaction between constituents, 
forming a black hole at the centre of the star would
generally require accumulating a very large DM mass.} There may be
model-dependent consequences of the energy released by inelastic DM
self-interactions, which in the case of fusions could release large
amounts of energy into small volumes over short timescales.


\section{Modifications to elastic scattering and direct detection}
\label{sec:elastic}


\subsection{Dark sector form factors}
\label{sec:ff}

For elastic scattering between a point-like state and
a spatially extended state, interacting via short-range interactions
sourced by some density $\rho$ on the extended side, the
 dependence on the spatial properties is summarised by
the form factor
\begin{equation}
F(\vecq) = \int d\vecr \, e^{i \vecq \cdot \vecr} \rho(\vecr) ,
\end{equation}
i.e.\ by the Fourier transform of the density.
For scattering of DM states off SM nuclei, effective field theory
arguments show that the scattering operator on the SM
side should take on one of a restricted number of forms (see e.g.~\cite{Fitzpatrick2012}
for a recent comprehensive analysis).
The dark sector density must have a complementary tensorial structure,
but could in general be determined by any properties
of the state, e.g. density, spin, etc.

In the case of two spatially extended states scattering off each other,
the effective density is the convolution of the separate spatial profiles, so the matrix
element is found by multiplying the form factors together.
Thus, the overall form factor for DM-SM scattering will be $F_N(\vecq) F_X(\vecq)$,
where $F_N$ and $F_X$ are, respectively, the SM nuclear form factor and the dark sector form
factor (as we will discuss below, in many experimental circumstances the SM nuclear
form factor is of limited importance).
The specific density determining the form factor will depend on the
nature of the DM-SM interaction. For simplicity, in this paper we will
restrict ourselves to considering the case of a scalar form factor
depending only on the number density of the state, e.g.\ arising from
the exchange of a heavy scalar mediator coupling uniformly to all
of the constituents. Taking, as discussed in the Introduction, the
the composite DM states to consist of an approximately uniform density
of constituent matter, a first approximation for the density is a spherical
top hat function, leading to a spherical Bessel function form factor,
\begin{equation}
F(q) = \frac{3 A j_1 (q R)}{q R}
= \frac{3  A  (\sin (q R) - q R \cos (q R))}{(q R)^3} ,
\label{eq:elasticff}
\end{equation}
where $R$ is the radius of the top hat density, and $A$ is the total
volume integral of the source across the distribution, so $F(0) = A$.

\begin{figure}
\begin{center}
\includegraphics[width=.6\textwidth]{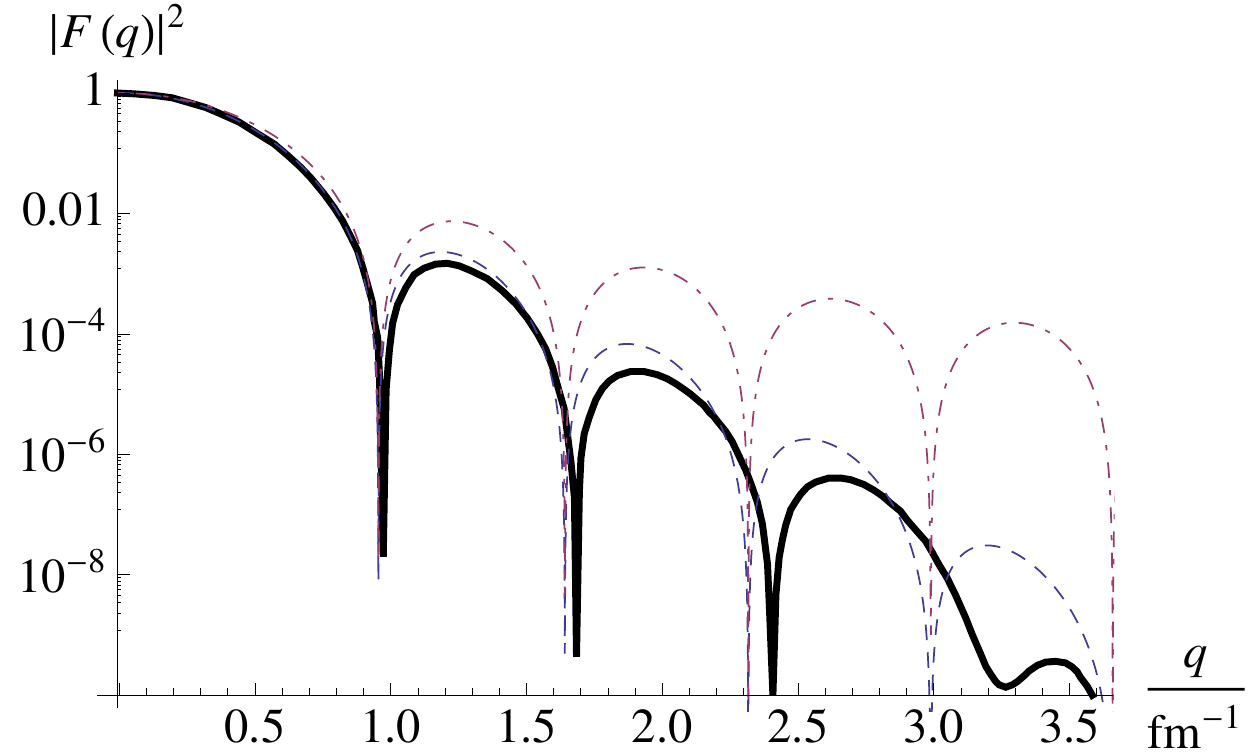}
\caption{Form factor for nuclear charge density (black) of $^{70}$Ge nucleus,
as derived from electron scattering data~\cite{Duda2006}.
Blue (dashed) curve is standard Helm parameterisation, red (dot-dashed) curve is
constant-density approximation.}
\label{fig:ff1}
\end{center}
\end{figure}

SM nuclei provide an example of this kind of roughly-constant-density
state, and illustrate the kind of deviations from the top-hat form
factor that might occur. Figure~\ref{fig:ff1} shows an example
of the form factor corresponding to the nuclear charge density
for the particular isotope $^{70}$Ge~\cite{Duda2006}, as inferred from
electron scattering data. Generally, and in this specific case, the first few peaks
and troughs of SM nuclear form factors are well-approximated by the `Helm' functional
form
\begin{equation}
F(q) = \frac{3 A j_1 (q R)}{q R} e^{-q^2 s^2 / 2}  ,
\label{eq:helmff}
\end{equation}
which is simply the top hat form factor modified to have a finite-width
fall-off, over the `skin depth' $s \simeq 0.9 \fermi$ (comparable to
the scale of the individual nucleons).

Importantly, in the case of interest to us where both a dark-sector and SM form-factor are present,
if the dark states have larger radii than SM nuclei, then the first
zeros, or more generally troughs, of $F_X(q)$ will occur at smaller $q$
than for $F_N(q)$. This means that, while the SM nuclear form factor is
of limited importance in most direct detection experiments, the dark
form factor may have interesting structure in precisely the
momentum,  and thus recoil energy, ranges being probed.


\subsection{Direct detection recap}
\label{sec:dd}

In this section, we briefly review the standard direct detection formalism
leading to the differential event rate (equation~\ref{eq:drder}),
mainly to establish notation for what follows. Readers familiar with this material may
safely proceed to Section~\ref{sec:coherence}.

The differential scattering rate (event rate per unit target mass)
at which incident DM particles scatter off an initially-stationary
target SM nucleus, giving it recoil energy $E_R$, is given by
\begin{equation}
\frac{dR}{dE_R} = \frac{1}{m_N} \int_{v > v_{\min}} d^3\vecv\, n_X f(\vecv) v \left.\frac{d\sigma_{XN}}{dE_R}\right|_v ,
\end{equation}
where $v \equiv |\vecv|$, $m_N$ is the mass of the SM nucleus, $n_X
f(\vecv) d^3\vecv$ is the differential number density of incident
DM particles, $v_{\min}$ is the minimum velocity required to obtain
recoil energy $E_R$, and $\sigma_{XN}$ is the DM-nucleus scattering
cross section.\footnote{Throughout we will consider parameter ranges where the DM-SM interactions are sufficiently weak that the Earth is optically thin to DM,
so that $n_X f(\vecv)$ is the DM halo distribution at the location of the Earth.} For non-relativistic elastic scattering, $v_{\min} =
\sqrt{\frac{E_R m_N}{2 \mu_{XN}^2}}$, where $\mu_{XN} = \frac{m_X
m_N}{m_X + m_N}$ is the DM-nucleus reduced mass.

The momentum transfer in an elastic collision is $q = \mu_{XN} v  \sqrt{2
(1-\cos\theta^*)}$, where $\theta^*$ is the scattering angle the
CoM frame. Since $E_R = \frac{q^2}{2 m_N}$, we have $dE_R = 
\frac{2\mu_{XN}^2 v^2}{m_N} \frac{d\Omega^*}{4 \pi}$,
where $d\Omega^*$ is differential solid angle in the CoM frame.
Referring back to the discussion of Section~\ref{sec:ff},
the matrix element for scattering at angle $\theta^*$ will
depend on the momentum transfer $\vecq$, and possibly also
on the velocity $\vecv$. Treating isotropic, velocity-independent scattering
to start with, and writing $\frac{1}{4\pi}\sigma_{XN}(q) \equiv
\frac{d\sigma}{d\Omega^*}$, we have
\begin{equation}
\left.\frac{d\sigma_{XN}}{dE_R}\right|_v = \frac{m_N}{2 \mu_{XN}^2 v^2} \sigma_{XN}(q) .
\end{equation}
So, under the assumption of velocity-independent scattering, we can factor
the differential scattering rate as 
\begin{equation}
\frac{dR}{dE_R} = \left(\int_{v > v_{\min}} d^3 \vecv\, \frac{f(\vecv)}{v} \right)
\frac{n_X}{2 \mu_{XN}^2} \sigma_{XN}(q)
\equiv g(v_{\min}) 
\frac{n_X }{2 \mu_{XN}^2} \sigma_{XN}(q) .
\label{eq:recoil1}
\end{equation}
Altering notation somewhat from Section~\ref{sec:ff},
we can write $\sigma_{XN}(q) = \sigma_{XN} F_N(q)^2 F_X(q)^2$,
where $\sigma_{XN}$ is the zero-momentum-transfer cross section (including
any coherence enhancement),
so that $|F_N(0)| = |F_X(0)| = 1$.

The scattering rate is usually expressed in terms of the DM scattering
cross section with nucleons, instead of with full nuclei. For dimension-6 interactions
between the DM constituents and SM quarks, the zero-momentum-transfer
scattering cross section goes as $|C|^2 \frac{\mu^2}{\Lambda^4}$,
where $\Lambda$ is the suppression
scale associated with the interaction,
and $C$ is the coherence enhancement factor.
Here, this is given by the product 
of the integrated densities relevant to the microscopic interaction
for the SM nucleus and the DM state, which we take
to be simply the respective nucleon numbers.
Thus
$\sigma_{XN} = \frac{|C_N|^2}{|C_n|^2} \frac{\mu_{XN}^2}{\mu_{Xn}^2} \sigma_{Xn}$,
where $\mu_{Xn}$ is the DM-nucleon reduced mass,
and so\footnote{This equation describes what a detector with perfect energy resolution would
see. If a detector has some response function $\kappa$ such that 
the differential rate to detect events with true energy $E_R$, occurring
at rate $R$, at measured energy $E_R'$, is $R \kappa (E_R, E_R') dE_R'$,
then the differential rate for measured events is
$R_d(E_R) = \int dE_R' \, \kappa (E_R, E_R') \frac{dR}{dE_R'}$.
}
\begin{equation}
\frac{dR}{dE_R} = g(v_{\min}) \frac{n_X}{2 \mu_{Xn}^2} \frac{|C_N|^2}{|C_n|^2} \sigma_{Xn} F_N(q)^2 F_X(q)^2 .
\label{eq:drder}
\end{equation}
%


\begin{figure}
\begin{center}
\includegraphics[width=.49\textwidth]{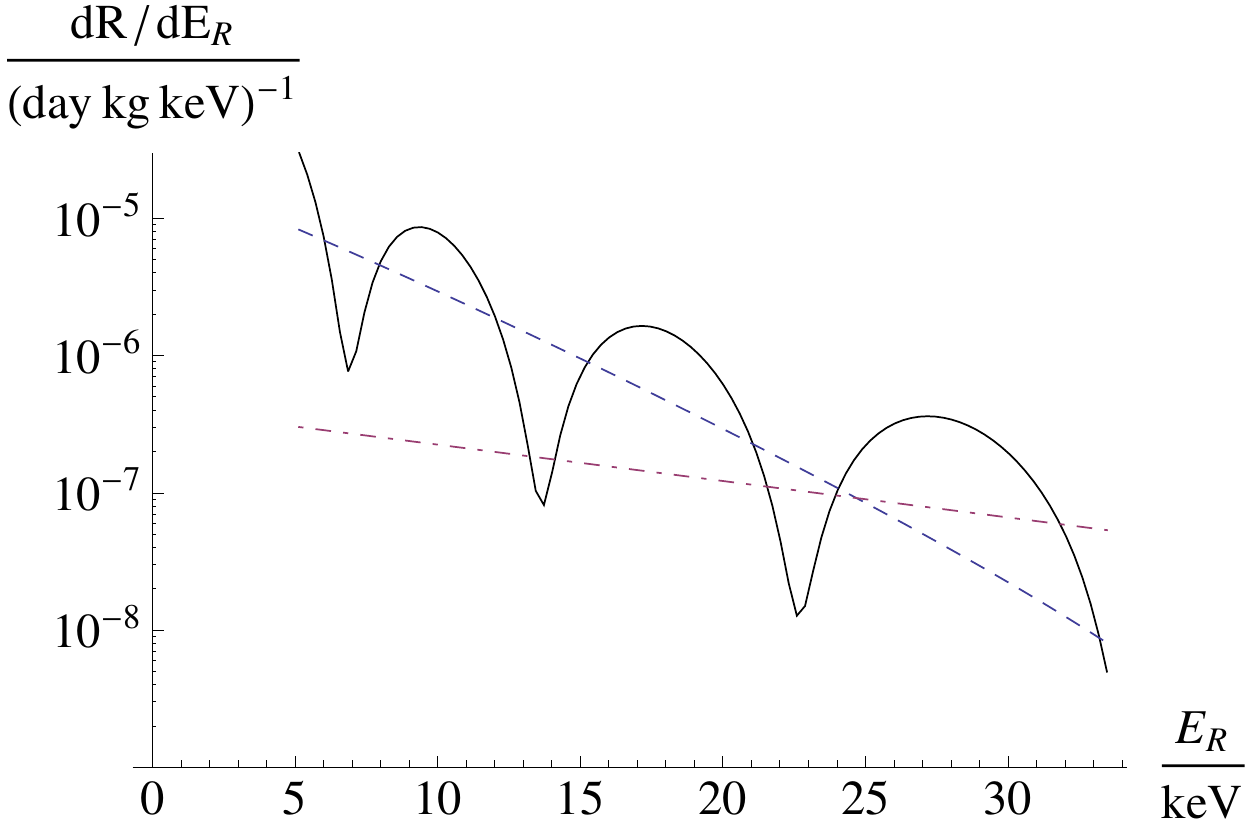}
\includegraphics[width=.49\textwidth]{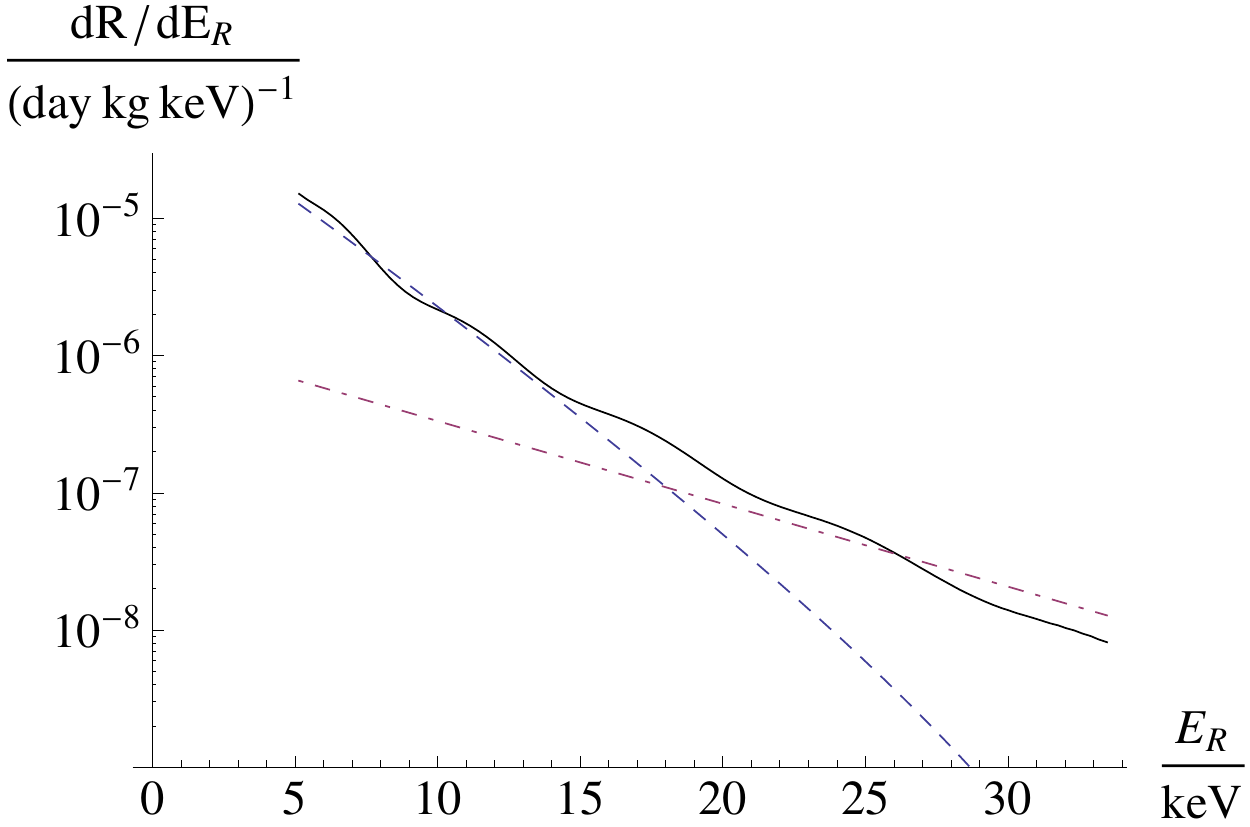}
\caption{Recoil energy spectrum (black, solid curves) for Bessel-function form factor,
radius $50 \fermi$, with DM velocity distribution as given by Standard
Halo Model~\cite{McCabe2010}, assuming \emph{Left:} Germanium detector target,
with Gaussian energy response ($\sigma(E) =
\sqrt{0.3^2 + 0.06^2 E / \keV} \keV$), and \emph{Right:} Xenon detector target,
with Gaussian energy response ($\sigma(E) = 0.6 \keV \sqrt{E/\keV}$),
assuming $5\keV$ energy threshold for both.
The DM state is taken to be composed of $3 \times 10^6$ constituents, each of
mass $20 \GeV$, with constituent-SM nucleon cross section of $2 \times 10^{-13} \pb$.
For comparison, blue (dashed) curves show energy recoil spectrum
for a $20 \GeV$ WIMP, red (dot-dashed) for a $1 \TeV$ WIMP,
both with $\sigma_{Xn} = 10^{-9} \pb$.
Energy response functions are taken from projections for future experiments
in~\cite{Cherry2014}.
}
\label{fig:besselRecoil}
\end{center}
\end{figure}


\subsection{Coherent enhancement of scattering rates}
\label{sec:coherence}

With the assumptions made in the previous section, the zero-momentum
transfer cross section between a DM state with $A$ constituents and a SM
nucleus with $N$ constituents is $\sigma_0 \sim A^2 N^2 \frac{\mu_{XN}^2}{\Lambda^4}$.
Taking the mass of the DM state to be $\propto A$, 
the DM number density is $\propto \frac{1}{m_{\rm X}}
\propto \frac{1}{A}$, assuming temporarily that all of the DM is of the same
size.  So, the overall scattering rate in direct detection experiments
will, for $m_{\rm X} \gg m_N$, be $\propto A$  for fixed
SM-constituent interactions. This is true if the DM radius $R$ is small
enough that typical momentum transfers do not probe $q R \gtrsim 1$
--- otherwise, the scattering rate is suppressed by the DM form factor,
as discussed in Section~\ref{sec:ff}. While, as discussed in Section~\ref{sec:sdist},
there might be some number distribution over DM states with different radii, if 
the mass distribution and scattering-rate distributions are confined
to a small range in logarithmic size, then the $\propto A$ enhancement will
hold approximately.

Figure~\ref{fig:besselRecoil} shows an example of the direct detection recoil energy
spectra resulting from a scenario of this kind, corresponding to
DM particles with a Bessel-function form factor of radius $50\fermi$.
These are compared to the recoil spectra for standard momentum-independent
(i.e.\ $F_X = 1$) WIMP scattering. Note that the constituent-SM nucleon
cross section required in the composite model is much smaller
than that required in WIMP models giving approximately the same event rates.
An additional point is that the much better energy resolution
expected in solid-state experiments would enable them to resolve the peaks
and troughs of a dark form factor corresponding to much larger radii
than for liquid-phase (e.g.\ Xenon) experiments.

The possible $\propto A$ enhancement means that, for a given
direct detection event rate, the expected production of lighter single
constituents in SM processes in colliders is reduced. For the example
of DM coupling through the Higgs portal, the strongest collider
constraint for DM states with $m_X < m_h/2$ comes from the Higgs invisible width.
As described in~\cite{Djouadi2011}, a bound of $< 10\%$ on the
invisible branching ratio of the Higgs puts constraints on the
DM-Higgs coupling a factor of a few better than current direct detection
experiments for DM masses a small factor lower than $m_h/2$. Thus, only
a modest relative suppression of collider rates vs direct detection
rates is needed to render the direct
detection bounds more constraining. Taking the example in
Figure~\ref{fig:besselRecoil}, with DM constituents of mass
$20 \GeV$ each having a nucleon scattering
cross section $\sigma_n = 2 \times 10^{-13} \pb$, we would require
Higgs invisible width bounds of $\sim 10^{-5}$ to be competitive with
current direct detection experiments.


\subsection{Recoil spectrum from size distribution}
\label{sec:sdist}

The previous sections considered a dark form factor arising
from DM particles having a single, common size. However, in many
models of large composite states (e.g.\ \cite{ndm1}, Q-balls),
the cosmological process through which a population of
these states arises generates a distribution over multiple sizes,
which would lead to a smeared-out signature in energy recoil spectra.

For DM consisting of a set of states $X_i$, the recoil
spectrum for elastic scattering will be
\begin{equation}
\frac{dR}{dE_R} = F_N(q)^2 \sum_i g_i(v_{\min}(\mu_{i,N})) \frac{n_i}{2 \mu_{i,n}^2}
\frac{|C_i|^2}{|C_n|^2} \sigma_{i,n} F_{X,i}(q)^2 ,
\end{equation}
where the subscripted quantities are those of equation~\ref{eq:drder}
for each species $X_i$.
In the specific case of a spectrum of related states, this may simplify somewhat.
Absent astrophysical self-interactions (see Section~\ref{sec:selfint}),
all of the states may be expected to have the same velocity distribution, i.e.\ $g_i = g$.
From $\sigma_n \sim |C|^2 \frac{\mu_{Xn}^2}{\Lambda^4}$, we expect the $\mu_{i,n}$
dependence to cancel, and additionally, if all 
of the states are heavy ($m_i \gg m_N$), then $\mu_{i,N} \simeq m_N$,
so we can factor out the $g$ dependent term fully.
Overall,
\begin{align}
\frac{dR}{dE_R} &\simeq \frac{n_X}{2 m_n^2} F_N(q)^2 g(v_{\min}) 
\frac{|C|^2}{|C_n|^2} \sigma_n
\sum_i \frac{n_i}{n_X} \frac{|C_i|^2}{|C|^2} F_{X,i}(q)^2 \\
&\equiv \frac{n_X}{2 m_n^2} F_N(q)^2 g(v_{\min}) \frac{|C|^2}{|C_n|^2} \sigma_n
F_X(q)^2 ,
\end{align}
where $n_X \equiv \sum_i n_i$,
and we have replaced the sum over individual states with a single effective
form factor, choosing $C$ appropriately so that $F_X(0) = 1$.

\begin{figure}
\begin{center}
\includegraphics[width=.49\textwidth]{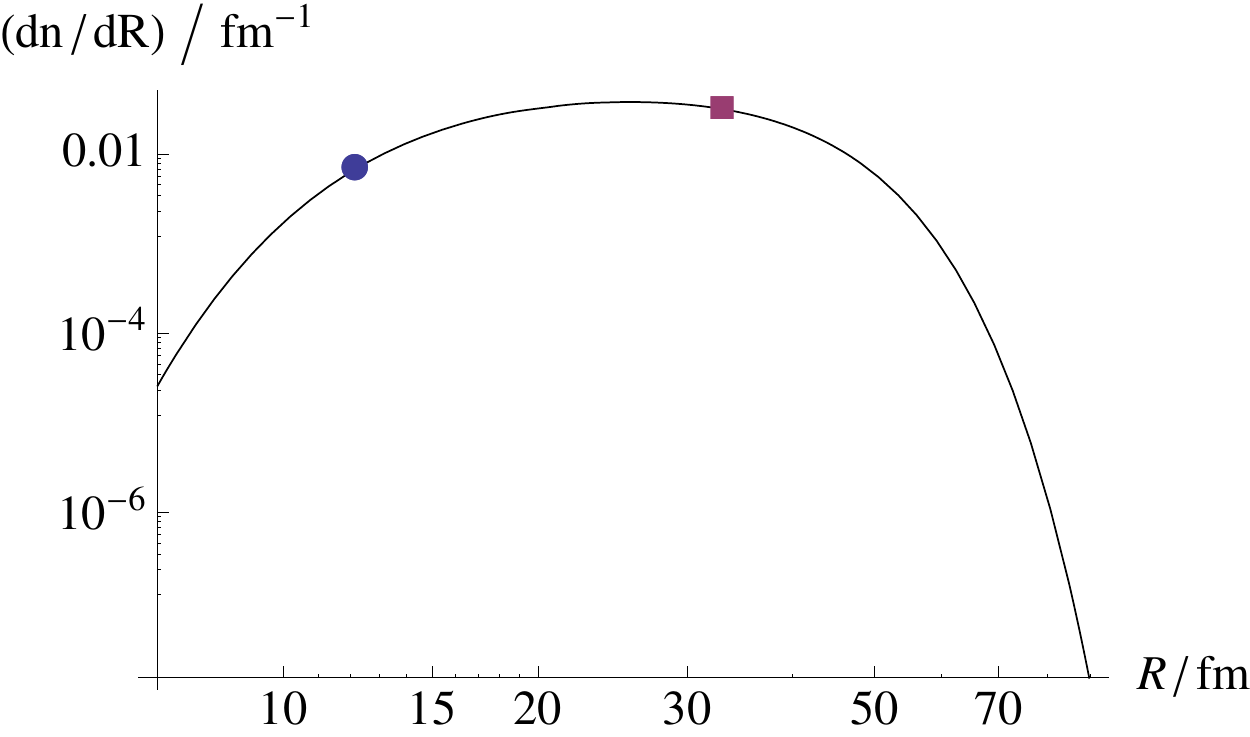}
\includegraphics[width=.49\textwidth]{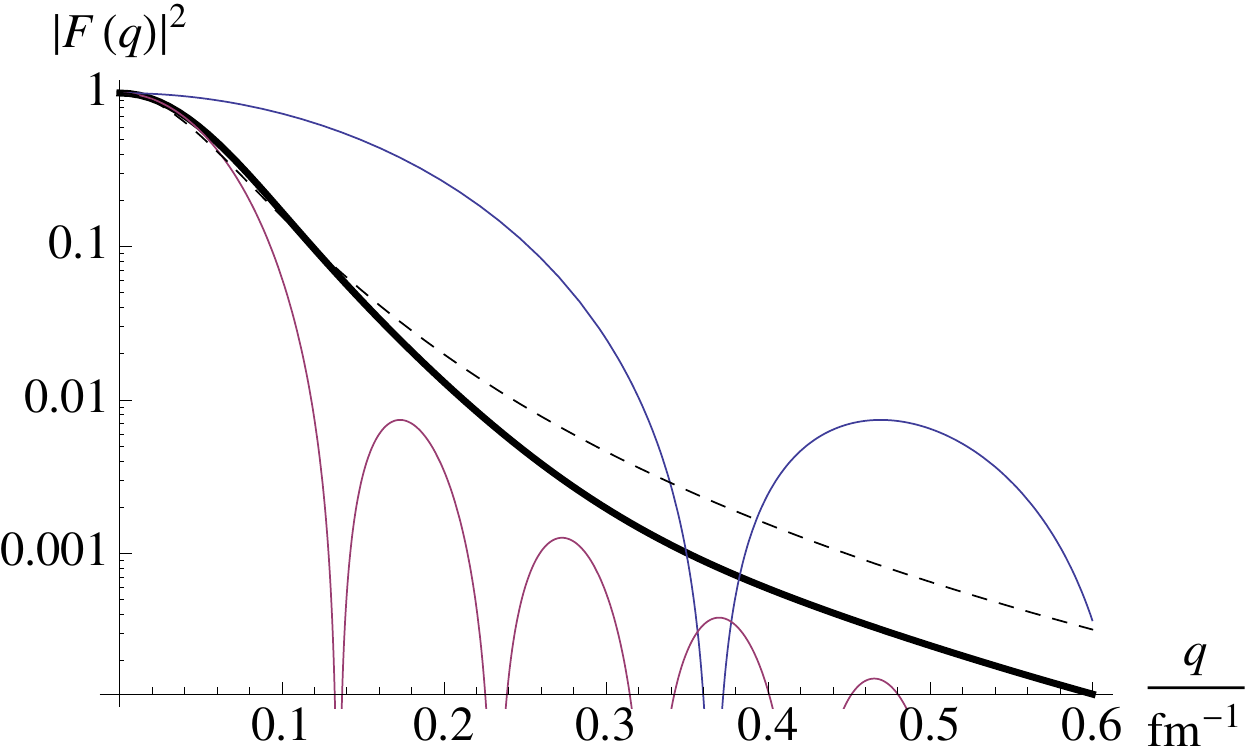}
\caption{\emph{Left:} Example (non-dimensionalised) number distribution of DM states
arising from nucleosynthesis-type process (from~\cite{ndm1}). 
\emph{Right:} form factors for DM sizes corresponding to 
points on left-hand plots (upper curve to blue circle, lower curve to red square),
and (black) effective form factor for whole number
distribution. Dashed line shows form factor for light
mediator particle, $F(q) \propto 1/(q^2 + m^2)$,
with $m = 16 \MeV$.}
\label{fig:ndist}
\end{center}
\end{figure}

Figure~\ref{fig:ndist} shows the effect of a distribution
over DM states of the kind considered in~\cite{ndm1}.
The sum over different sizes corresponds to adding together
differently scaled form factors, which smooths out the troughs of the Bessel-function
form factor. In this case, we end up with a form factor which is similar
to that obtained from the exchange of a light
mediator particle, $F(q) \propto \frac{1}{q^2 + m_\phi^2}$,
but arising entirely from contact interactions, without the need
for a light state.\footnote{
While it is possible that a model with an intermediate-mass
mediator would have additional phenomenology distinguishing it from
the case of large composite DM, this would not have to be the case.
Though a light mediator will give rise to self-interactions between
galactic DM particles, for heavy (and thus dilute) DM these will not
be frequent enough to have detectable effects on halo shapes.
The direct detection event rate is set by the products of the
squared couplings for the SM and the dark sector, so by making
the dark sector coupling large, we could make the required SM coupling
very small. Current direct detection experiments are sensitive enough that
any future signals would imply a minimum value of the SM coupling
small enough to be significantly below any direct production constraints.
Of course, the SM coupling may be above this minimum if the dark sector
coupling is also small, so direct production signals are not ruled out.
Also, as discussed in~\cite{ndm1}, direct bounds on the
SM couplings of light ($m \ll 100 \MeV$) states generally imply
that they cannot lose any initial cosmological energy density they have
to the SM sufficiently fast (except possibly to neutrinos), requiring
the introduction of additional light hidden sector states that persist
to the present day.
}
For
a given DM velocity distribution,
these kinds of dark form factor gives an energy recoil spectrum with
a shape different from that of standard momentum-independent scattering,
falling off more slowly at high energies than a low-mass WIMP, and more quickly
at low energies than a higher-mass one. 
In the next section, we discuss how such form factors could be distinguished
from momentum-independent scattering.

\subsection{Dependence on DM velocity distribution}
\label{sec:veldist}

From the definition in equation~\ref{eq:recoil1} of $g(v_{\min}) = \int_{v > v_{\min}}
d^3\vecv \frac{f(\vecv)}{v}$, it is clear that, by choosing $f(\vecv)$ appropriately,
we can make $g$ any non-increasing function of $v_{\min}$
(with the physical constraint that it must fall off very fast
past $v_e + v_{\rm esc}$, where $v_e$ is the velocity of Earth relative
to the Galactic rest frame, and $v_{\rm esc}$ is the Galactic
escape velocity at the position of the Earth). 
Although the Standard Halo Model, which posits a Maxwell-Boltzmann DM velocity
distribution, is commonly assumed, it is highly plausible
that the DM velocity distribution at Earth differs from this,
perhaps significantly.\footnote{Simulations of dark halo formation
indicate that non-Maxwell-Boltzmann velocity distributions may
generically arise~\cite{Kuhlen2010}, and also suggest the possibility
of a co-rotating DM disc~\cite{Read2009},
along with sub-structure including cold streams~\cite{Diemand2008}.}

This has the consequence that, for elastic scattering, the energy
recoil spectrum can be anything of the form $F_N(q)^2 F_X(q)^2$ multiplied
by a non-increasing distribution, as discussed in~\cite{Fox:2010bu}.
So, even though the energy recoil
spectrum for the dark form factor looked different from that
for a WIMP \emph{assuming the same DM velocity distribution},
by changing the velocity distribution we could make a WIMP mimic
the (averaged) form factor spectrum, since this is decreasing with $E_R$.

However, if we have data from multiple experiments, each of which
uses a different type of SM target nucleus, this degeneracy can be lifted.
On the assumption that there is no dark form factor, the event
distributions (once we have compensated for the different target
nuclei) with respect to $v_{\min}$ should be the same for each of the
experiments \cite{Fox:2010bz,Fox:2010bu,Frandsen:2011gi,Bozorgnia:2013hsa,DelNobile:2013cva,Scopel:2014kba} --- any disagreement indicates the presence of some extra
effect. For a dark form factor, since $q(v_{\min}) = 2\mu_{XN} v_{\min}$, changing $\mu_{XN}$
by changing $m_N$ will change the range of $F_X(q)$ that we sample (significantly
so if the DM mass is larger than those of the SM target nuclei).
Ref~\cite{Cherry2014} performs a multi-target analysis along these
lines for various DM models, and a forthcoming paper~\cite{CherryNew}
applies this method to the case of heavy DM coupling via an intermediate-mass mediator,
which as noted above gives form factors that are very similar to
those which may be obtained via averaging over Bessel-function form
factors. It is found\footnote{We thank John Cherry for providing plots of the exposure time needed to distinguish our effective form factor.} that such a form factor can be distinguished from
contact-like scattering with only a handful of events in multiple
detectors, while effective exposures of only around 1.0 ton years in multiple
detectors could, assuming DM-SM cross sections close to current upper bounds, allow
the determination of the mediator mass (which, in the composite case,
would correspond to the average size of the DM) to around 25\% accuracy.


\subsection{Detectability of a rising energy recoil spectrum}
\label{sec:rising}

While Sections~\ref{sec:sdist} and~\ref{sec:veldist} have considered the
situation of a distribution over DM sizes, the more striking scenario
of DM states being concentrated around a single size, giving
rise to peaks and troughs in the energy recoil spectrum as per
Figure~\ref{fig:besselRecoil}, has qualitatively different detectability
in direct detection experiments.
As described in the previous section, the energy recoil spectrum for
elastic scattering can be anything of the form $F(q)^2$ multiplied
by a non-increasing distribution. While this allows non-increasing
sections of an energy recoil spectrum to be explained by a combination
of form factor and DM velocity distribution, any rises in the spectrum
must come from the form factor (or from the presence
of inelastic scattering). Though
summing over multiple Bessel-function widths will generally smooth
out the individual peaks into a falling distribution, if
the particle sizes were clustered mostly around a
particular value,\footnote{
For example, due to the binding energy per constituent reaching
a maximum value at some size, and decreasing after that (as for iron
in the SM), rendering larger states unstable to fission.} 
then the troughs/peaks of such an effective form
factor could give a rising recoil spectrum.
Assuming a detector has sufficiently
good energy resolution so as not to smooth out the troughs in the
underlying recoil spectrum, it will be possible, with sufficiently many events,
to rule out point-like elastic scattering.
Furthermore, if both the falling and rising parts of a trough were visible,
this would be a clear sign of a more complicated momentum-dependent form factor,
or of some combination of inelastic scattering modes.

\begin{figure}
\begin{center}
\includegraphics[width=.49\textwidth]{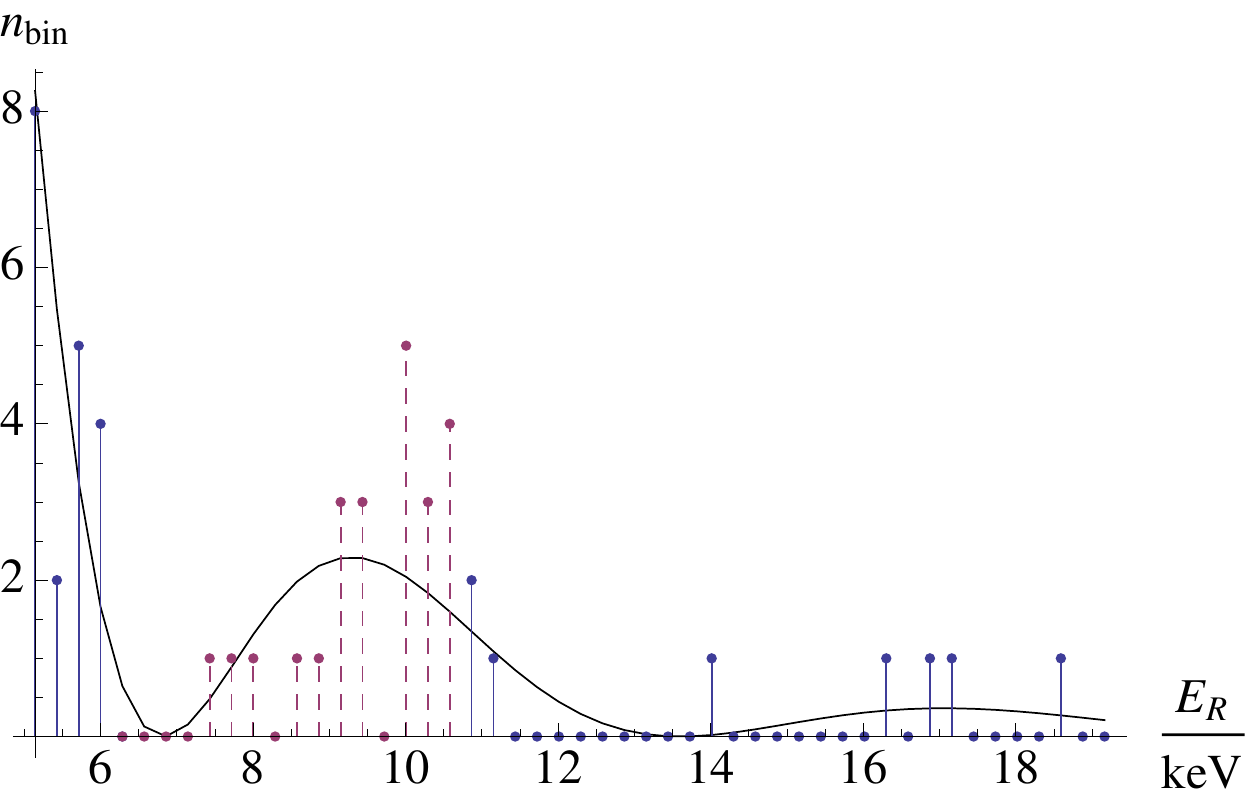}
\includegraphics[width=.49\textwidth]{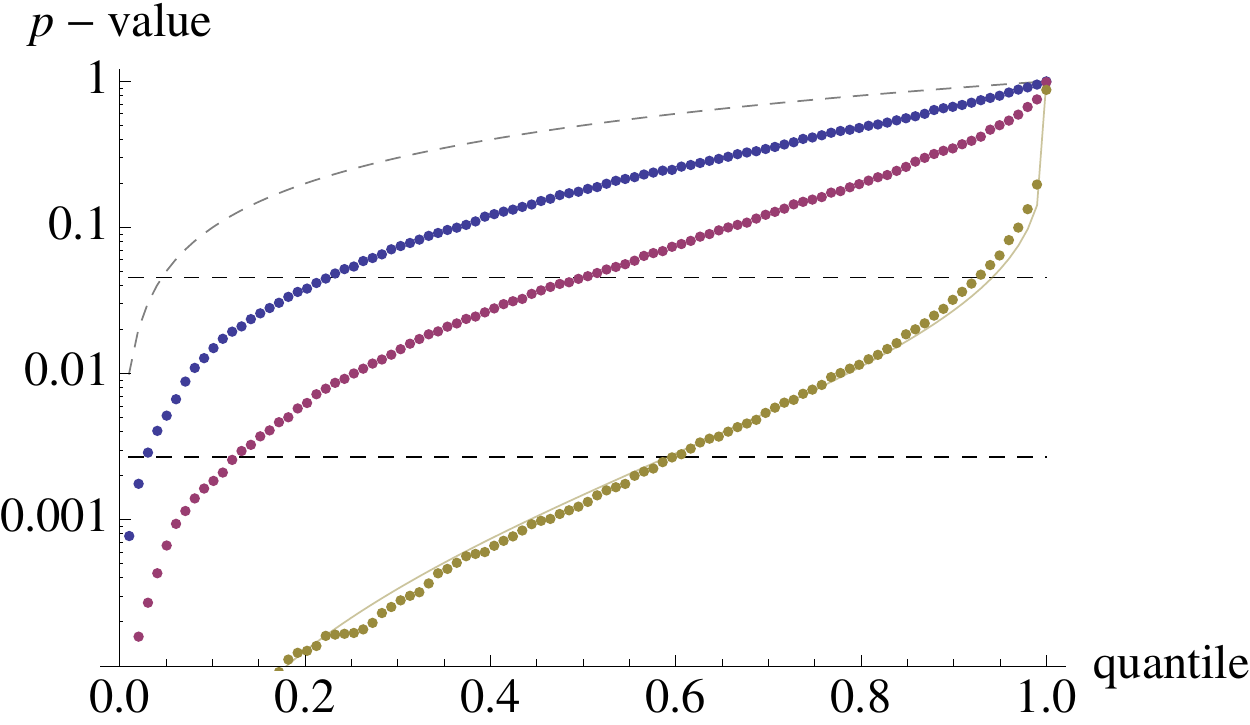}
\caption{\emph{Left:} Energy recoil spectrum (black) for a Bessel-function
form factor corresponding to radius $50 \fermi$, with
other parameters as per left-hand plot of Figure~\ref{fig:besselRecoil}.
Blue points (solid lines) are a particular sample of 50 events from
this distribution, binned with an energy width of $0.3\keV$. Red points (dashed lines) show the
interval of points that is worst fit by a non-increasing distribution
(see text). The $p$-value for the test described in
Appendix~\ref{app:rising} is $0.005$.
\emph{Right:} Cumulative distribution functions (CDFs) for $p$-value from test described in Appendix~\ref{app:rising},
for 30 samples (blue), 50 samples (red), and 100 samples (yellow).
Grey (dashed) curve shows the CDF for a uniform underlying distribution. The
upper and lower dashed lines shows the $p$ values corresponding to
$2\sigma$ and $3\sigma$ significance for rejecting the hypothesis of
a non-increasing distribution.
}
\label{fig:fallingp}
\end{center}
\end{figure}

We can test whether a given
set of events came from a non-increasing distribution by locating the
energy interval with the `worst' bias towards its high-energy end ---
here measured simply by the average energy of the events --- and
asking what the probability is that a given sample from a candidate
non-increasing distribution would have an interval that extreme
(explained in more detail in Appendix~\ref{app:rising}).
Figure~\ref{fig:fallingp} displays the results of applying this test to
simulated data from a model with a Bessel-function dark form factor,
scaled so that a number of the peaks/troughs are visible in the energy
recoil spectrum. As illustrated, around 12 expected events in the
most significant rising section, corresponding here to $\sim 50$
events overall, are sufficient to obtain a two-sigma exclusion of the
non-increasing hypothesis in the majority of cases.
While a realistic analysis would need to take into account the distribution
of background events and other issues, this illustrates that relatively
few events from only a single detector may suffice to give very interesting
physical information about DM properties.


\section{Low-energy excitations \& inelastic scattering}
\label{sec:inelastic}


\subsection{Properties of low-energy excitations}
\label{sec:excitations}

Another feature generic to large composite states
is the presence of low-lying modes, leading to
the possibility of inelastic scattering in which these
modes are excited.  Such modes can be broadly
characterised as either of ``single-particle'' or ``collective'' type.
We are most interested in the collective modes since, as we will discuss, their
excitation amplitudes can be coherently enhanced compared
to the single-particle excitations, and thus can dominate the inelastic scattering
rate if they are of low enough energy to be excited.

There are a number of forms such collective excitations could take.
The simplest and most generic type of collective excitations
are simply vibrational modes of either the bulk or of the surface
of the large composite state.
Alternatively, if the constituents have spins, moments, or other intrinsic
properties, then collective oscillations of these properties, such as spin waves,
can occur.
In addition, if the ground state is not spherically
symmetric, so the composite state is deformed in shape as happens for large
SM nuclei, then there will be low-lying rotational modes as well.
For definiteness, in this paper we will focus on the inelastic scattering involving the most
universal and model-independent of these collective modes, namely, the surface and bulk
vibrational modes that are of long wavelength
compared to the scale of the individual constituents.

For large composite states consisting of roughly uniform `nuclear
matter', the lowest-energy density waves will generally be surface
waves, i.e.\ volume-preserving oscillations of the `nuclear
surface'~\cite{BohrMottelson}. The reason for this is that the energy of
the lowest bulk compressional excitations is set
by the speed of sound for such waves, $c_c \sim \sqrt{\frac{K}{\rho}}$, where $K$ parameterises
the compressibility of the nuclear material, and $\rho$ is the density.   Taking
into account that the lowest possible wavenumber is $k \sim 1/R$, the energy of the low-lying
compressional excitations is given by
\begin{equation}
\delta E_c \sim \sqrt{\frac{K}{\rho}} \frac{1}{R}~.
\end{equation}
On the other hand, the speed of volume-preserving surface capillary waves is set
by $c_s \sim \sqrt{\frac{\sigma k}{\rho}}$, where $\sigma$ is the surface tension
(this also holds in relativistic hydrodynamics~\cite{Coleman1985}), so 
the energy of the low-lying surface excitations is given by
\begin{equation}
\delta E_s \sim \sqrt{\frac{\sigma}{\rho}} \frac{1}{R^{3/2}}~.
\end{equation}
Using the facts that $R \sim A^{1/3} R_1$, where $R_1$ is the length scale of a single constituent, and that both
the surface tension and bulk compressibility are set by the same underlying interaction strength between
the basic constituents so we have $\sigma \sim R_1 K$, leads to 
\begin{equation}
\frac{\delta E_s}{\delta E_c} \sim  A^{-1/6}~.
\end{equation}
Thus for large enough $A$ the energy of the bulk compressional modes is well separated from the
lower-lying surface modes. 

We now summarise the dynamics of the surface modes more quantitatively.
Classically, for small-amplitude surface waves of a homogeneous, 
incompressible, sharp-edged fluid droplet,
we have $\rho(r,\theta,\phi)$ constant inside $R'(\theta,\phi,t)$ and zero outside, with
\begin{equation}
R'(\theta,\phi,t) = R \left[ 1 + \sum_{l \ge 2} \sum_{m=-l}^l \alpha_{lm}(t) Y_{lm}(\theta,\phi)\right] ,
\label{eq:rprime}
\end{equation}
for amplitude coefficients $\alpha_{lm}(t)$.  The $l=1$ modes are removed as they correspond to
oscillations of the CoM position, while the $l=0$ mode is removed as it is just the monopole compression oscillation.\footnote{If the amplitude of the surface waves were not small then volume-preserving
oscillations require $\alpha_{00} = - \sum_{l\ge2,m} |\alpha_{lm}|^2$, but this constraint can be ignored at leading order.}
The total Hamiltonian of the surface excitations is
\begin{equation}
H = \frac{1}{2} \sum_{\substack{l\geq 2,\\ m=-l,...,+l}} 
\left( B_l \left| \dot\alpha_{lm}(t) \right|^2 + 
C_l \left| \alpha_{lm}(t) \right|^2 \right) ,
\end{equation}
where $B_l = \rho R^5/l$ and $C_l=(l-1)(l+2) R^2 \sigma$ are the
``mass'' and ``stiffness'' parameters of each mode.

Quantising this system leads to each mode being an independent quantum harmonic oscillator.
The mode frequencies, $\omega_{lm}$, of the oscillations are independent of the azimuthal spherical harmonic
parameter $m$ and are given by~\cite{Coleman1985}
\begin{equation}
\omega_{l} = \sqrt{\frac{C_l}{B_l}} = \left( \frac{l(l+2)(l-1)\sigma}{R^3 \rho}\right)^{1/2}.
\end{equation}
The overall excitation energy spectrum
\(\Delta E_{\{ n_l \}} = \sum_{l\geq 2} n_l \omega_l\)
is thus set by the
occupation numbers $n_{l}$ of the $(2l+1)$-fold-degenerate $l$-modes.
It will be important for our discussion of the coherent excitation
probability of these modes that the simple harmonic oscillator-like
wavefunctions associated to each mode are characterised by a length
scale, $R\epsilon_l$, where the dimensionless parameter $\epsilon_l$ is
given by
\begin{equation}
\epsilon_l = \frac{1}{\sqrt{2 B_l \omega_l}}~.
\label{eq:epsl}
\end{equation}
(For an individual mode with occupation number $n_l\gg 1$ the corresponding classical amplitude of oscillation
simply scales as $|\alpha_l| \sim \epsilon_l \sqrt{n_l}$.)
In terms of the underlying parameters of the composite state
\begin{equation}
\epsilon_l \propto A^{-7/12} \left(\frac{1}{M_1 R_1^2 \beta}\right)^{1/4}~,
\label{eq:epslscaling}
\end{equation}
where $\beta\sim \sigma R_1^2$ is the energy associated to surface tension, and $M_1$
is the mass of individual constituents.
As discussed in~\cite{Tassie1962}, the hydrodynamic values of $B_l$ and $C_l$
can differ by factors of $\mathcal{O}(10)$ from those obtained experimentally
for SM nuclei, due to the effects of shell structure etc. However, the hydrodynamic
approximation should give a good qualitative guide to the properties of the low-lying
modes of large composite states.

Finally, as mentioned above, as well as collective modes there may also be
modes corresponding to the excitation of `single constituents'.
If the interior of the state
consists of degenerate fermionic matter, then excitations which move a
state from just below the Fermi surface, through a small change in momentum $\vecq$
approximately tangential to the surface, will look like quasi-particles of energy
$\vecq^2 / (2 m_*)$, where the quasi-particle mass $m_*$ will often be of
order the constituent mass \cite{Mahaux:1985zz}. Since the smallest allowed
momentum change is $q \sim 1/R$, the lowest-energy quasi-particle
excitation will have energy $\omega \sim \frac{1}{2 R^2 m_\star} \sim A^{-1/3}/R$,
so for large $A$ will be of lower energy than the collective
surface modes. However, as discussed in the next section,
in the regimes of interest in this paper, the cross sections
for exciting them are much lower. Additionally, the properties of these
modes are model-dependent --- for example, pairing interactions between
constituents could lead to superfluidity, as arises in forms of SM nuclear matter.


\subsection{Inelastic scattering form factors}
\label{sec:inelasticff}

The presence of these low-lying modes means that even low-velocity
scattering processes may have sufficient energy to be inelastic. In
particular, scattering off the long-wavelength collective modes, since these involve
all of the constituents, will be coherently enhanced as for elastic
scattering, and have their own momentum-dependent form factors.

Generally, this form factor can be calculated from the overlap
of initial and final state wavefunctions. As derived in
Appendix~\ref{app:inelastic}, if the scattering occurs through
a scalar contact interaction with the constituents of the DM state,
then (returning to the notation of Section~\ref{sec:ff}) the form factor for inelastic scattering into a one-phonon
surface mode of angular momentum number $l$ is, as given
in equation~\ref{eq:surf1},
\begin{equation}
F(q) = 
\frac{3 A}{\sqrt{4 \pi}} i^l (2l+1)^{1/2} \epsilon_l
j_l (q R)
\end{equation}
(as compared to equation~\ref{eq:elasticff} for elastic scattering),
where $\epsilon_l$ is the natural amplitude associated with oscillations
in that mode. This result is correct
to first order in $\epsilon_l$, and valid for $q R \lesssim 1/\epsilon_l$
(beyond that, the wavefunction overlap can be computed numerically).
Excitations of multiple phonons are associated with further factors
of $\epsilon_l$. 
Comparing elastic to inelastic scattering, we see that the latter
has form factors corresponding to higher spherical Bessel functions,
and is suppressed by powers of the natural amplitude of the
surface modes; as per equation~\ref{eq:epslscaling}, this
is parametrically small for large $A$.

The cross sections for scattering off single-particle excitations
add incoherently, giving $\sigma \propto A$. There may also be
further suppression factors. The case of degenerate fermionic
matter may be well-approximated by scattering off a Fermi
gas of non-interacting quasi-particles --- as discussed in
Appendix~\ref{app:vlarge}, this is, for low momentum
transfers, strongly suppressed by degeneracy factors.
Generally, incoherent scattering should only be significant for momentum
transfers comparable to the Fermi momentum (and energy transfers comparable
to the Fermi energy), or for very large composite states where scattering from
collective modes is highly suppressed (Appendix~\ref{app:vlarge}).


\subsection{Inelastic recoil spectra}
\label{sec:inelasticrecoil}

If collisions are inelastic --- say, the DM state
is excited from $m_X$ to $m_X + \delta$ --- then
the minimum velocity~\cite{TuckerSmith:2001hy}
for which we can obtain a given recoil energy is~\footnote{
Equation~\ref{eq:vminin} is accurate up to multiplicative
correction of size at most $\sim \sqrt{\delta / M}$ --- this
always gives a good approximation, since 
$\delta < \frac{1}{2}\mu_{XN} v_{\max}^2$ is required for any excitations to occur,
and galactic escape velocity is deeply non-relativistic.}
\begin{equation}
v_{\min} = \frac{1}{\sqrt{2 m_N E_R}}\left(\frac{m_N E_R}{\mu_{XN}} + \delta\right)~.
\label{eq:vminin}
\end{equation}
We have $dE_R = \frac{2 p p'}{m_N} \frac{d\Omega^*}{4 \pi}$, where
$p$ and $p'$ are the initial and final momentum in the CoM frame.
The final state phase space available is also proportional to $p'$,
so the differential rates for elastic and inelastic scattering
are related by
\begin{equation}
\frac{d\sigma_i}{dE_R} = \frac{d\sigma_e}{dE_R} \frac{|\mathcal{M}_i|^2}
{|\mathcal{M}_e|^2}~,
\end{equation}
where the $\mathcal{M}$ are the matrix elements.
Aside from differences in the momentum-dependence of the matrix elements,
giving different form factors, the main qualitative difference from elastic
scattering is that $v_{\min}$ is no longer monotonically increasing with $E_R$,
instead having a minimum at $E_R = \delta \mu_{XN} / m_N$.
This is in contrast to elastic scattering, where the fact that
that $v_{\min}$ is monotonically increasing had
important consequences (see Sections~\ref{sec:veldist} and~\ref{sec:rising}) for data analysis.

As per above, we generically expect large
composite states to have low-energy excitations, which may be excited by
scattering with SM nuclei. The energy recoil spectrum will then be a sum
over the spectra for scattering into
each of these states (including the ground state, which
gives elastic scattering). Labelling the states by $\lambda$,
\begin{equation}
\frac{dR}{dE_R} = \frac{n_X}{2 \mu_{Xn}^2} \sigma_{Xn} F_N(q)^2 \sum_\lambda
g(v_{\min,\lambda}(E_R)) \frac{|C_\lambda|^2}{|C_n|^2} F_{X,\lambda}(q)^2 ,
\end{equation}
where $\frac{C_\lambda}{C_n}$ gives the ratio of matrix elements
(c.f.\ equation~\ref{eq:drder}), and the $F_{X,\lambda}$
have a common normalisation such that elastic scattering has $F(0) = 1$.

\begin{figure}
\begin{center}
\includegraphics[width=.6\textwidth]{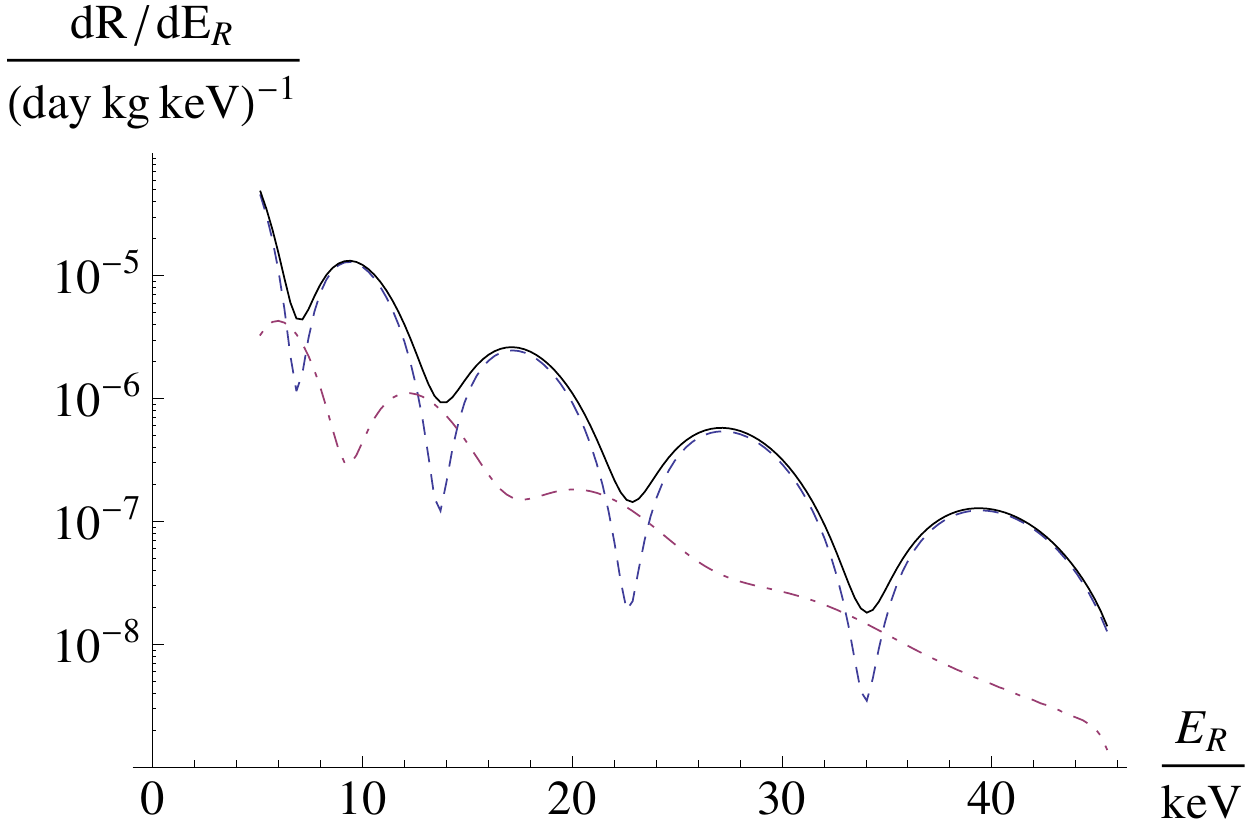}
\caption{Energy recoil spectrum (Germanium target) for spherical DM state with 
parameters as per Figure~\ref{fig:besselRecoil},
incorporating elastic scattering and inelastic excitations of surface
modes (black). Blue (dashed) and red (dot-dashed) curves show
contributions from elastic and inelastic scattering respectively. The
energy of the first surface mode is taken to be $28 \keV$, so a number
of modes contribute at higher
recoil energies, flattening out the inelastic recoil spectrum.
The amplitude of the inelastic modes is set a factor $\sim 10$
above the hydrodynamic estimate from Section~\ref{sec:excitations},
for illustrative purposes.
Note that the form factor for the first excited mode is 90 degrees
out of phase from that for elastic scattering --- this effect can be
seen in inelastic SM-SM scattering, e.g.\ Figure 1 of~\cite{Tassie1962}.}
\label{fig:inelastic1}
\end{center}
\end{figure}

Figure~\ref{fig:inelastic1} shows an example of the energy recoil
spectrum arising from a uniform-density composite state
of the kind discussed in Section~\ref{sec:excitations},
in the regime where elastic scattering, and inelastic scattering
into the first few surface modes, are the dominant effects.
Due to the relative suppression of inelastic modes, from both
form factor effects and the small amplitude of surface oscillations, elastic scattering
generally dominates in such scenarios.

Finally we comment that the above analysis assumed that the DM particles are
dominantly in the ground state.  If de-excitation times are long enough, there is
a possibility that there may be a significant cosmological population of excited DM states,
altering the inelastic scattering phenomenology with exothermic interactions
being possible.  We leave discussion of this highly model-dependent possibility for future
work.


\section{Astrophysical capture}
\label{sec:capture}

In addition to potentially being visible in direct detection experiments,
interactions between DM and SM matter may also lead to the capture of DM
by astrophysical objects, the most interesting generally being various
kinds of stars. Once enough DM has accumulated inside the star, then
either self-interactions among the DM (annihilations, self-scattering,
etc.), or more complicated DM-SM interactions, may in some models lead to
observable alterations of stellar properties. For definiteness,
we consider the case of asymmetric DM, in which DM may build up
in the star without being destroyed through annihilations.

For spatially-extended DM states, the possibility of large cross
sections for inelastic DM-DM interactions may give rise to modifications
of the distribution of captured DM inside the star, analogously to
the possibility of effects on astrophysical halo shapes mentioned
in~\cite{ndm1}. In models with large velocity kicks after exothermic
collisions, these could take the form of ejecting captured DM from the
star, as discussed in e.g.~\cite{Detmold2014}. Alternatively, other
models could lead to the contraction of the captured DM distribution,
potentially all of the way down
to a very dense configuration at the centre of the star. Since 
we have been considering composite states
composed of similar constituents, some kind of
repulsive self-interaction countering the attractive binding forces
is already required in such models, and will generally remove the
danger of such a DM configuration collapsing into a
black hole.\footnote{As may occur in models of very heavy fundamental asymmetric DM
\cite{Zurek2013,Kouvaris2012}} 
In the absence of any further dynamics, there are no obvious
externally observable consequences of a high-DM-density central
region. In particular, the release of binding energy from
fusions will have no significant effect on the star---for symmetric
DM, almost all of the DM captured by the star can generally annihilate with no observable
heating, and we are only injecting a small fraction of this energy.
However, in the case of run-away fusions, there may be inelastic
collisions between extremely large composite states, which could
result in the release of large amounts of energy, in small volumes and
on rapid timescales. If some of these de-excitation products couple 
to the SM strongly enough, the SM energy injection resulting from these
processes may have consequences --- for example, \cite{Graham:2015apa} points
out that sufficiently fast and localised energy injection inside a white dwarf
may ignite a Type 1a supernova, even in a sub-Chandrasekhar mass dwarf.


\subsection{Effect of dark form factors on capture rate}

The composite nature of the DM may also affect the initial capture rate,
through the momentum-dependence of scattering cross sections.
From~\cite{Gould1987}, the capture rate per unit volume is given by
\begin{equation}
\frac{dC}{dV} = \int_0^\infty du\, \frac{f(u)}{u} w \, \Omega (w)~,
\end{equation}
where $u$ is the velocity at infinity of the DM particle relative to the Sun,
$f(u)$ is the DM speed distribution at infinity, $w$ is the velocity of the DM
particle at the scattering location, and $\Omega(w)$ is the scattering rate
to less than escape velocity off the material at that location.
The scattering rate is
\begin{equation}
\Omega(w)  = \sum_i n_i w \, \Theta\left(E_{\max,i} - E_{\min}\right) 
\int_{E_{\min}}^{E_{\max,i}} dE \, \frac{d\sigma_i}{dE}~,
\end{equation}
where $n_i$ is the number density of SM nuclear species $i$, $E_{\min} = \frac{1}{2} m_X u^2$ is the minimum energy loss 
required for capture, and $E_{\max,i} = \frac{2 \mu_i^2}{m_i} w^2$ with $\mu_i \equiv m_i m_X / (m_i + m_X)$
is the maximum kinematically-allowed energy loss (taking the `cold sun' approximation
where the scattering nuclei are at rest). 
In the case where the scattering involves momentum dependent form
factors, then using the notation of Section~\ref{sec:dd}, 
\begin{equation}
\Omega(w)  = \sum_i n_i \frac{\sigma_{Xi}}{2 \mu_i^2 w} \, \Theta\left(q_{\max,i} - q_{\min,i}\right)
\int_{q_{\min,i}}^{q_{\max,i}} q dq \, F_i(q)^2 F_X(q)^2~,
\end{equation}
where $q_{\min,i} = \sqrt{m_i m_X} u$, and $q_{\max,i} = 2 \mu_i w$.

Combining into a single form factor $F(q) \equiv F_i(q) F_X(q)$,
if $F(q) \simeq 1$ for $q < 1/R$, then for $q_{\max,i} R = 2 \mu_i w R \lesssim 1$ 
the form factor will have no significant effect. If $q_{\max,i} R \gg 1$,
but $q_{\min,i} R \lesssim 1$, then assuming that $F(q)^2$ falls
off faster than $1/q^2$,\footnote{In particular, the Fourier transform of a  spherically symmetric distribution with
at worst step function discontinuities eventually falls off at least as fast
as $1/q^2$.} the
dominant contribution to the integral will be from momenta $\sim 1/R$,
giving
\begin{equation}
\int_{q_{\min,i}}^{q_{\max,i}} q dq \, F(q)^2 
\sim \frac{1}{R^2}~.
\end{equation}
If $q_{\min,i} R \gg 1$ as well, then the integral is 
\begin{equation}
\int_{q_{\min,i}}^{q_{\max,i}} q dq \, F(q)^2
\sim q_{\min,i}^2 F(q_{\min,i})^2 \Lambda(q_{\min,i},q_{\max,i})
\end{equation}
where $\Lambda$ will be $\sim 1$, if $q_{\min,i} \ll q_{\max,i}$ and
$F(q)^2$ does not fall off significantly faster than a power law in $q$
(if $q_{\min,i} \ge q_{\max,i}$, then $\Lambda = 0$).
Turning to the integral over DM halo velocities, if the velocity distribution
is steeply falling past some typical velocity $\bar{u}$ and approximately constant
(in $d^3 u$ space) below that, as for a Maxwell-Boltzmann distribution,
then $f(u) \simeq n_X u^2 / \bar{u}^3$, and
we can approximate the integral as 
\begin{equation}
\frac{dC}{dV} \simeq \frac{n_X}{\bar{u}^3} \int_0^{\bar{u}} u \, du \, w \Omega(w)~.
\end{equation}
If $q_{\max,i}(\bar{u}) R = 2 \mu_i w(\bar{u}) R \lesssim 1$, then no part
of this integral has any significant momentum dependence
(for capture off species $i$).
If $q_{\max,i} R \gg 1$ throughout (i.e. $2 \mu_i v_{\rm esc} R \gg 1$, where $v_{\rm esc}$ is the escape velocity), but $q_{\min,i}(\bar{u}) R
= \sqrt{m_i m_X} \, \bar{u} R \lesssim 1$, then the integral becomes
\begin{equation}
\int_0^{\bar{u}} u \, du \, w \Omega(w) \simeq 
n_i \frac{\sigma_{Xi}}{2 \mu_i^2} \frac{1}{R^2} \frac{\bar{u}^2}{2}~.
\end{equation}
For momentum-independent scattering with $q_{\max,i} \gg q_{\min,i}(\bar{u})$,
\begin{equation}
\int_0^{\bar{u}} u \, du \, w \Omega(w) \simeq 
n_i \frac{\sigma_{Xi}}{2 \mu_i^2} 4 \mu_i^2 v_{\rm esc}^2 \frac{\bar{u}^2}{2}~,
\label{eq:momindep}
\end{equation}
(where we assume that $v_{\rm esc} \gg \bar{u}$, so that terms of the
form $\bar{u}^4$ can be ignored), so the suppression relative to this
case is $\frac{1}{4 \mu_i^2 v_{\rm esc}^2 R^2} \equiv \frac{1}{(k R)^2}$,
where $k/2 = \mu_i v_{\rm esc}$ is approximately the initial momentum in the CoM
frame.

If $q_{\min,i}(\bar{u})R \gg 1$, then
\begin{multline}
\int_0^{\bar{u}} u \, du \,w \Omega (w) 
\simeq
n_i \frac{\sigma_{Xi}}{2 \mu_i^2} \left(
\int_0^{1/(\sqrt{m_i m_X} R)} u \, du \, \frac{1}{R^2}
\, + \right. \\
\left. \int_{1/(\sqrt{m_i m_X} R)}^{\bar{u}} u \, du \, m_i m_X u^2 F(\sqrt{m_i m_X} u)^2 
\Lambda (\sqrt{m_i m_X} u, 2 \mu_i w)
\right).
\label{eq:udu1}
\end{multline}
If $F(q)^2$ drops off faster than $1/q^4$, then this integral is dominated
by $u \sim 1/(\sqrt{m_i m_X} R)$, giving 
\begin{equation}
\sim n_i \frac{\sigma_{Xi}}{2 \mu_i^2} \frac{1}{2 m_i m_X R^4} 
= 
n_i \frac{\sigma_{Xi}}{2 \mu_i^2} \frac{1}{R^2} \frac{\bar{u}^2}{2}
\left(\frac{1}{m_i m_X R^2 \bar{u}^2}\right),
\end{equation}
which has an extra suppression factor. If, as in the Bessel-function
case, $F(q)^2 \propto 1/q^4$ for large $q$, then the second integral
in equation~\ref{eq:udu1} contributes some logarithmic multiple of
$1/(m_i m_X R^4)$.

To summarise, for heavy DM ($m_X \gg m_i$) with a radius of $R$
large enough to be probed in collisions, $k R \gg 1$,
the elastic capture rate off species $i$ will be suppressed relative
to momentum-independent scattering by
$\sim (k R)^{-2}\min\left(1, \frac{\zeta}{m_i m_X R^2
\bar{u}^2}\right)$, replacing the usual kinematic suppression\footnote{
For momentum-independent scattering, if $q_{\min,i}(\bar{u}) > q_{\max,i}$,
then the integral in equation~\ref{eq:momindep} is cut off above when $q_{\min,i}(u) \simeq q_{\max,i}$, i.e.\ when $u \sim \sqrt{m_i/m_X} v_{\rm esc} \equiv u_c$ (taking $\mu_i \simeq m_i$).
This reduces the value of the integral by a factor of $\sim u_c^2 / \bar{u}^2
= m_i v_{\rm esc}^2/(m_X \bar{u}^2)$.
}
$\min\left(1,\frac{m_i v_{\rm esc}^2}{m_X \bar{u}^2}\right)$
(here, $\zeta$ is some logarithmic integral).
As an example, if we suppose that the Sun (escape velocity $0.002 c$) has $k R \simeq 1$,
then capture in white dwarfs ($v_{\rm esc} \sim 0.02 c$) is suppressed
by a factor of $\sim 10^{-3}$,
and in neutron stars ($v_{\rm esc} \sim 0.7 c$) by
$\sim 10^{-5}$,
compared to standard elastic scattering (for white dwarfs, the SM
nuclear form factor is also important). 
These suppressions mean that inelastic scattering may be the dominant
capture process, though we leave the investigation of this to future work.


\subsection{Self-interactions of captured dark matter}
\label{sec:selfint}

One possible effect of large DM self-interaction cross sections
is to increase the capture rate, in that incoming DM can scatter off
already-captured DM as well as off SM nuclei in the star.
However, for heavy DM, and in locations of reasonable astrophysical
DM density, it is generally hard to accumulate enough
captured DM to have any significant effect on the overall capture rate.
Supposing that the star has effective capture cross section $A_s$
for DM particles streaming through it,
the total number of DM particles it captures
in its lifetime is $N_X \sim A_s t_s v n_X$,
where $n_X$ is the local DM number density and $v$ is
the characteristic relative velocity (of order galactic
orbital velocities, $\sim 220 \km \sec^{-1}$).
Writing the DM self-scattering cross-section as $\sigma_{XX}$,
the total cross sectional area for DM-DM scattering is
at most
\begin{align}
A_X &= \sigma_{XX} N_X \sim (\sigma_{XX} / m_X) t_s v \rho_X A_s \\
&\simeq A_s \left(\frac{\sigma_{XX} / m_X}{\barn / \GeV}\right) \frac{v}{220 \km \sec^{-1}} \frac{t_s}{5 \Gyr} \frac{\rho_X}{0.3 \GeV \cm^{-3}}~.
\label{eq:ax}
\end{align}
For elastically-scattering DM with velocity-independent self-scattering
cross section, the observational limit on the self-scattering
cross section is $\sigma_{XX} / m_X \lesssim 1 \barn / \GeV$ (see e.g.~\cite{Tulin2013}).
That $A_X$ can be of order $A_s$ for this value means that
there are possible DM models in which self-capture is significant
(e.g.~\cite{Zentner2009,Bertone2010}). However, for large composite DM states,
the fact that $\sigma_{XX} / m_X \propto m_X^{-1/3}$ for uniform matter means
that large states generally stand little chance of having significant self-capture.
Evaluating $\sigma_{XX} / m_X$ for a composite state of interior density $\rho_b$,
and constituent number $A$,
\begin{equation}
\frac{\sigma_{AA}}{m_A} \simeq \frac{0.05 \barn}{\GeV} 
A^{-1/3}
\left(\frac{1 \GeV}{m_1}\right)^{1/3} 
\left(\frac{1 \GeV \fermi^{-3}} {\rho_b}\right)^{2/3} ,
\end{equation}
we see that for SM-like or heavier constituents, large composite
states are well below elastic self-interaction bounds, so do not
give interesting self-capture effects.

However, interactions between already-captured DM states could
potentially have interesting effects. Particularly interesting are
interactions which may lead to `run-away' effects, in which 
the DM distribution inside the star contracts to a very dense
state. This could arise either from inelastic collisions which
are dissipative (i.e.\ some of the initial KE is lost into de-excitation
products), or from fusions, which result in heavier DM states
that have a correspondingly smaller equilibrium radius inside
the star. In Sections~\ref{sec:endo} and~\ref{sec:fusions},
we will perform some approximate calculations to demonstrate
the feasibility of these scenarios.


\subsubsection{Dissipative collisions}
\label{sec:endo}

For dissipative collisions, the dynamics are governed by
the rate at which DM-DM collisions dissipate energy,
versus the rate at which DM-SM collisions add it, re-thermalising the DM.
As an estimate, if some fraction $\alpha$ of integrated phase space density
for DM lies within a spatial radius $r$, then 
the rate of self-interactions is set by 
\begin{equation}
\Gamma_{XX} \sim \sigma_{XX}
n_X v_X \sim \frac{\sigma_{XX}}{r^2} \alpha N_X  \frac{3}{4 \pi}
\sqrt{\frac{G M_\star}{R_\star^3}}
\sim 10 \dy^{-1} \alpha \frac{A_X}{r^2}~,
\end{equation}
where we have taken $v_X^2 \sim \frac{G M(r)}{r}$,
and $A_X$ is as per equation~\ref{eq:ax}.
In thermal equilibrium with the SM matter in the star,
the DM states would have an isothermal distribution,
\begin{equation}
\rho_X \sim e^{-r^2/(2 r_*^2)} \quad , \quad
r_* = \left(\frac{3 T_\star}{4 \pi G m_X \rho_\star}\right)^{1/2}~,
\end{equation}
where $T_\star$ and $\rho_\star$ are the temperature and density of the
stellar core.
Evaluating $\Gamma_{XX}$ for $r = r_*$, and putting in values appropriate
to the Sun, 
\begin{equation}
\Gamma_{XX} \sim 6 \yr^{-1} \alpha 
\left(\frac{\sigma_{Xn} / m_X}{10^{-8} \pb / \TeV}\right)
\left(\frac{m_X}{\TeV}\right)^{2/3} \left(\frac{1 \GeV \fermi^{-3}}{\rho_b}\right)^{2/3}~,
\label{eq:gxx}
\end{equation}
where we have used equation~\ref{eq:csun} for the solar capture rate,
and we have replaced $\sigma_{XX}$ by the geometrical cross
section between two spheres of mass $m_X$ and internal density $\rho_b$.\footnote{A
$1\TeV$ mass sphere of this density has radius $\simeq 6 \fermi$, which
is of the order of the inverse momentum transfer in the initial DM-nucleus collisions
(for solar capture), so there will not be significant momentum suppression
of the capture rate.}
This interaction timescale is much less than the lifetime of the Sun.
So, as long as scattering with SM particles does not counteract the
increase in the local DM number density that comes about from losing energy
in inelastic scatterings, it is inconsistent for a large fraction
of the captured DM to have a basically isotropic steady state phase space distribution
with most of the density within a sphere of isothermal or smaller radius.
Thus, if the DM-SM cross section is large enough, the DM distribution
may contract down to a very dense configuration at the centre of the
star.

It remains to check that SM scatterings do not re-thermalise DM fast enough to
avoid this contraction.
The viscous drag force on a DM state is approximately
$\vecf_{\rm drag} \sim - m_N n_N v_N \sigma_{XN} \vecv_X
\equiv -\gamma m_X \vecv_X$,
where $m_N$, $n_N$ and $v_N$ are the mass, number density and
average speed of the SM scatterers, $\sigma_{XN}$
is the momentum transfer cross section between $X$ and the scatterers,
and $\vecv_X$ is the velocity of $X$ in the thermal rest frame
of the scatterers. Thus, $\gamma$ gives the damping rate
at which scatterings return $\vecv_X$ to the thermal distribution.
For parameters relevant to the Solar core,
\begin{equation}
\gamma \simeq \frac{m_N n_N v_N \sigma_{XN}}{m_X}
\simeq 3 \times 10^{-4} \yr^{-1}
\left(\frac{\sigma_{Xn} / m_X}{10^{-8} \pb / \TeV}\right)~.
\label{eq:gamma1}
\end{equation}
Thus, as long as each inelastic collision does not dissipate
too small a fraction of the initial KE ($\sim 10^{-4}$ for the parameters
in equations~\ref{eq:gxx} and~\ref{eq:gamma1}), SM-DM thermalisation
will be too slow to prevent contraction. Also, since the thermalisation timescale
is much less than the lifetime of the star, we do expect to contract down
to isothermal densities in the first place (in particular, this means
that we will lose almost all of any net angular momentum the captured DM distribution
might have had initially, meaning that rotational support of the collapsing
distribution will not be a worry).
The same separation between self-interaction rates and thermalisation times
may, for suitable parameter ranges, apply in the case of white dwarfs
and neutron stars as well (however, for neutron stars there is an issue
of whether thermalisation timescales are less than the lifetime of the
star~\cite{Bertoni2013}).


\subsubsection{Fusions}
\label{sec:fusions}

An alternative way to realise run-away contraction is for
the DM states to progressively increase in mass through fusions,
and correspondingly for their isothermal radius within the star to decrease.
In contrast to the previous section, 
this scenario relies on thermalisation through SM-DM scattering
being \emph{fast} enough. It also relies on fusion cross sections being large
enough, and on fusions remaining energetically favourable
up to very large sizes --- these features may naturally be realised in
models of composite DM such as those discussed in~\cite{ndm1}.

As discussed in Appendix~\ref{app:vlarge}, we expect the coherent momentum
transfer cross section between DM and SM states to scale as
$\sigma_{AN} \propto A^{2/3}$ for states $A$ with $p R_A \gg 1$,
where $p$ is the characteristic momentum of the SM scatterers.
Thus, $\gamma_A \simeq (A_p/A)^{1/3} \gamma_{A_p}$ for $A \gg A_p$,
where $A_p$ is the size for which the DM states are of radius $\sim 1/p$,
i.e.\ $p R_{A_p} \sim 1$, 
with $\gamma$ given by equation~\ref{eq:gamma1}.
In terms of the size $A_i$ of states which have radius $\sim 1/p_i$, where
$p_i$ is the characteristic momentum of SM scatterers in the \emph{initial}
collision (which occurs at around the escape velocity),
thermalisation is naively effective (assuming small enough injections
of kinetic energy on the DM side) for
$A \lesssim 10^{26} A_i \left( \frac{\sigma_{A_i n} / m_{A_i}}{10^{-8} \pb / \TeV}\right)^3\equiv A_{\rm th}$. 
The isothermal distribution for such heavy states may already
have volume comparable to their saturated volume. If not, then
further fusions will generally be fast enough to combine most of the DM
into a few very large composites --- the initial number density, and so
rate of fusions, will be very high, 
with build-up continuing as $\Gamma_{AA} \propto A^{-1} A^{2/3}
\sim A^{-1/3}$ from number density and cross section factors respectively.

For white dwarfs and neutron stars, we have a similar conclusion
that fusions are potentially fast enough to form a very dense configuration.
In these cases, the star is compact enough that even a DM distribution
of radius comparable to the entire star can still have a high enough
self-interaction rate for fusions to combine the majority of the DM
into a dense state (though we would still need to worry about dissipating
energy and angular momentum, especially in the neutron star case).

We note that the estimates in Section~\ref{sec:endo} and~\ref{sec:fusions}
should be viewed as rough plausibility estimates --- proper
investigation of these issues would
require realistic modelling of the DM-DM collisions, and of the phase
space distribution of DM at each stage of the process. In particular,
this section has ignored the possibility that DM collision types other
than fusions are important (e.g. fragmentations etc.), and also assumed
that the velocity kicks imparted by fusion de-excitation products are
small enough to be re-thermalised quickly. However, the point was merely
to illustrate that run-away contraction is a plausible possibility.


\section{Conclusions}

In this paper, we have investigated some of the consequences that follow
if a proportion of DM is composed of large composite
states --- specifically, states consisting of a large number of constituents
forming an extended, semi-uniform object with a saturated density.

The spatial extension of these objects introduces a dark form factor
into elastic scattering amplitudes as discussed in Section~\ref{sec:elastic}. 
This has two effects:  First, coherently enhanced scattering rates with the result
that collider bounds on DM cross sections are effectively weakened by factors of up to $1/A$
(Section~\ref{sec:coherence}).  Second, the introduction of an additional characteristic 
momentum dependence, whose effects on direct detection
energy recoil spectra, for DM states with radii large compared to SM nuclei, are
considered in Sections~\ref{sec:coherence} and~\ref{sec:sdist}.
We find that, in the most visible cases, such signals may be distinguishable
from elastic scattering, in a halo-velocity-independent manner,
with only $\mathcal{O}(50)$ scattering events (Section~\ref{sec:rising}).

Large composite states will generically have long-wavelength collective
excitations, which give rise to the possibility of low-energy,
coherently enhanced inelastic scattering processes, as discussed in
Section~\ref{sec:inelastic}. Though such processes will generally
be sub-dominant in direct detection experiments (Section~\ref{sec:inelasticrecoil}),
in specific models they may have signatures, in direct detection or astrophysically.

We considered the capture of DM by stars in Section~\ref{sec:capture}.
The presence of dark form factors and inelastic scattering
may alter the capture rate.  In addition we investigated the possibility of
inelastic interactions between captured DM states having an effect on the DM distribution
inside the star.  Strikingly, for plausible types of self-interactions,
we argue that it is possible for almost all of the captured DM to accumulate
into a very dense configuration (Sections~\ref{sec:endo} and~\ref{sec:fusions}).

Finally, we emphasise that our calculations have focused on what we believe to be the leading, 
most model-independent features associated with the interaction of large composite DM states with
the SM sector.  In specific realisations of large composite DM there may well be other even more striking
signals.  It would clearly be interesting to construct specific models which realise large composite
DM and investigate the resulting phenomenology.


\section*{Acknowledgements}

We wish to thank Asimina Arvanitaki, Mads Frandsen, Peter Graham, Felix Kahlhoefer, Surjeet Rajendran and Ian Shoemaker for useful discussions, and Joan Elias Miro for useful comments on a draft. We are particularly grateful to John Cherry for detailed discussions of exposure times. RL and JMR thank the Stanford Institute for Theoretical Physics and the Perimeter Institute
for Theoretical Physics, and SMW thanks the Oxford Physics Department,
for hospitality during completion of this work.
RL acknowledges support from an STFC studentship.


\appendix

\section{Matrix elements for inelastic scattering}
\label{app:inelastic}

Suppose that we have a scattering process in which a state $i$, with
initial momentum $\veck$, scatters off a state $I$ with initial momentum $\vecp$
(all non-relativistic), resulting in final states $f$ and $F$ with momenta
$\veck'$ and $\vecp'$ respectively. The matrix element for this, in the Born
approximation, is
\begin{align}
\mathcal{A} &= 
\langle f, \veck'; F, \vecp' | H_{\rm int} | i, \veck; I, \vecp \rangle
\\
&=
\langle f, \veck; F, \vecp | e^{i \vecq \cdot (\hat{\vecx}_F - \hat{\vecx}_r)}
H_{\rm int} | i, \veck; I, \vecp \rangle
\\
&\equiv
\int d\vecr \, e^{i \vecq \cdot \vecr} V_{\rm eff} (\vecr, \vecv)~,
\end{align}
where $\veck' - \veck = \vecp - \vecp' \equiv \vecq$ is the momentum
transfer, $\vecv$ is the relative velocity of $i$ and $I$, and $\langle
\vecx_f; \vecx_F | H_{\rm int} | \veck; \vecp \rangle \equiv (2\pi)^{-3} V_{\rm
eff} (\vecx_f - \vecx_F, \vecv) e^{i \veck \cdot \vecx_f} e^{i \vecp
\cdot \vecx_F}$. That is, the matrix element is
given, as a function of the momentum transfer $\vecq$, by the Fourier
transform of a (possibly velocity-dependent) effective potential
$V_{\rm eff}$, whose form will depend on the interaction Hamiltonian.
If the initial or final states have directional properties (polarisations),
$V_{\rm eff}$ may also depend on those.

As discussed in Section~\ref{sec:excitations}, the surface modes of an
incompressible liquid drop can be obtained by quantising the classical
surface oscillations, in which the surface is displaced as in
equation~\ref{eq:rprime}. Classically, for a scalar interaction
between a plane-wave scatterer and the nuclear matter, the Fourier
transform of the interaction potential is given by
\begin{align}
F(\alpha_{lm}) &= \int_{r < R'(\theta,\phi)} d^3\vecr \, e^{i \vecq \cdot \vecr} V_0 \\
&= V_0 \int d\Omega \int_0^{R'(\theta,\phi)} r^2 dr\, 
\sum_{l'} (2l' + 1) i^{l'} j_{l'} (q r) P_{l'}(\cos\theta)~,
\label{eq:falpha}
\end{align}
where we have taken $\vecq$ to be in the $\vecz$ direction.
For $m \neq 0$, this integral is clearly zero.
Expanding to first order in a given $\alpha_{l0}$, and eliding $V_0$,
\begin{align}
F(\alpha_{l0}) &= F(0) + \sqrt{4 \pi} \int d\Omega \, \alpha_{l0} R^3 \, Y_{l0}(\theta) \sum_{l'} (2l' + 1)^{1/2} i^{l'} j_{l'} (q R) Y_{l'0}(\theta)
\\
&= F(0) + \alpha_{l0} \frac{3 A}{\sqrt{4 \pi}} (2l + 1)^{1/2} i^l j_l (q R)~.
\end{align}
Treating each mode as a harmonic oscillator, we have $\hat\alpha_{l0} = \epsilon_l (\hat a_l + \hat a_l^\dagger)$, where $\epsilon_l$ is the dimensionless amplitude
from equation~\ref{eq:epsl}, and $\hat a_l^\dagger$ is the creation
operator for the $l,0$ mode. So the matrix element between
the ground state and the first excited state is, to first order
in $\epsilon_l$, given by
\begin{equation}
\langle 1_l | F(\hat \alpha) | 0 \rangle = \frac{3 A}{\sqrt{4 \pi}} (2l + 1)^{1/2} i^l \epsilon_l j_l (q R)~.
\label{eq:surf1}
\end{equation}
Higher phonon number states are associated with further factors
of $\epsilon_l$ (for the appropriate $l$ numbers), and more factors
of $q R$ in front of the spherical Bessel function. For the expansion
to make sense, we must have that $j_l (q r)$ varies slowly over the interval
$R (1 \pm \epsilon)$, so we need $q R \lesssim \pi/\epsilon_l$.
Beyond this approximation, 
\begin{equation}
\langle 1_l | F(\hat \alpha) | 0 \rangle = \int d\alpha \,
\psi_1^*(\alpha) \psi_0(\alpha) F(\alpha)~,
\label{eq:surf2}
\end{equation}
where the $\psi_n$ are the oscillator wavefunctions
(for a harmonic oscillator, Hermite polynomials multiplied by an exponential),
and $F(\alpha)$ is from equation~\ref{eq:falpha}
(note that the oscillator properties derived in Section~\ref{sec:excitations}
are also to first order in $\epsilon_l$).
In particular,  comparing equation~\ref{eq:surf1} to the elastic scattering
form factor $3 A j_1 (q R) / (q R)$ from equation~\ref{eq:elasticff}, 
both are $\sim A \epsilon_l^2$ at $q R \sim 1/\epsilon_l$, but
the correct form of the inelastic form factor from equation~\ref{eq:surf2}
drops off faster than the elastic form factor beyond this,
since the wavefunction for the surface displacement integrates
over multiple, cancelling, Bessel-function periods.


\section{High-momentum scattering from large composite states}
\label{app:vlarge}

In this Appendix, we consider the scattering behaviour 
of large composite states when the momentum $p$ of the incoming scatterer
is such that $p R \gg 1$, where $R$ is the radius of the composite state.

\subsection{Coherent scattering}

The elastic scattering form factor should fall off at least as fast as $(q R)^{-2}$.
For thermalisation-type processes, we are most interested in the rate at
which SM scatterings exchange energy-momentum with the composite state,
which for dominantly soft scattering will be set parametrically
by the momentum transfer cross section $\sigma_{\rm tr} 
= \int (1 - \cos\theta) \frac{d \sigma}{d\Omega} d\Omega$.
Writing $\frac{d \sigma}{d\Omega} = \frac{\sigma_0}{4 \pi} f(q R) / (q R)^4$,
and using $q^2 = 2 \mu^2 v^2 (1 - \cos\theta)$ in the CoM frame,
\begin{equation}
\sigma_{\rm tr} = \frac{1}{2} \sigma_0 \int d\cos\theta (1 - \cos\theta) \frac{f(q R)}{(q R)^4}
= \frac{1}{2} \frac{\sigma_0}{(\mu v R)^4} \int_0^{2 \mu v} \frac{dq}{q} f(q R)~.
\end{equation}
The dimensionless integral on the RHS provides a logarithmic factor $\Lambda$, giving
\begin{equation}
\sigma_{\rm tr} \sim 
\frac{\Lambda}{(\mu v R)^4} \sigma_0 
= \frac{\Lambda}{(k R)^4} \sigma_0~.
\end{equation}
Since $\sigma_0 \propto A^2$, we have $\sigma_{\rm tr} \sim A^{2/3}$
for $A$ large enough that $k R \gg 1$ (with $\Lambda$ changing
only logarithmically with $k R$), as used in Section~\ref{sec:fusions}.

The above assumed that the form factor was that appropriate for plane-wave
scattering. If both the composite state and the SM scatterer are better-localised
than the size of the composite state, then the wavefunction overlap
in the form factor will not probe the full composite state, 
giving different results.

For inelastic scattering, taking the composite state to be much heavier and much larger
than the scattering state, we can approximate the scattering
as being against an infinite uniform medium (assuming that $p$ is small
enough that it does not resolve structure on the scale
of individual constituents). Since the collective modes are linearly
dispersing, $\omega = c_c k$, then by the usual pseudo-momentum and energy conservation
considerations, scatterers can only excite these if their velocity relative
to the medium is greater than $c_c$ (as per superfluids). Putting in
some illustrative numbers, the speed of CNO nuclei inside the Sun is $v
\simeq 3 \times 10^{-4} c$, while the speed of sound for compressional
modes of SM-like nuclear matter is $\sim 0.1c$.

For a composite state of finite radius, this corresponds to the fact that
the modes of small enough energy to be excited have $k \ll q$, so the wavefunction
overlap in the form factor is very small (the $q$ wave
oscillates much faster than the $k$ wave, giving a large cancellation).
With collective surface modes, as explained
in Appendix~\ref{app:inelastic}, for $\epsilon_l q R \gg 1$ there is
again a large cancellation.

The above comments assume that the composite state is in its ground
state. If there is some non-zero occupation number for high-wavenumber
modes, then down-scatterings of these are energetically permitted, and
will not suffer from the cancellation suppression described. We do not 
consider the case of scattering against excited states here,
assuming that the de-excitation times are much shorter than the times
between composite-SM collisions (for this to be the case, de-excitation will
generally have to be to hidden sector states).


\subsection{Incoherent scattering}

As discussed in Section~\ref{sec:excitations}, as well as the collective
modes, there may also be low-lying modes corresponding
to the excitation of `single constituents' --- in the
case of degenerate fermionic matter, excitations in which particles
just below the Fermi surface are scattered to just above it.
By the
usual Fermi-liquid theory, ignoring interactions and approximating the
scattering as occurring from a non-interacting Fermi gas should provide a
good first approximation. \cite{Bertoni2013} discusses scattering from
a degenerate Fermi gas, finding that the scattering rate for 
low-energy scatterers is independent of the Fermi momentum (as can be
seen from geometrical considerations).
This can result in large suppression factors compared to
the naive scattering rate given by $\sigma n v$ --- for scatterers
with momentum $k \ll p_F$, and also low velocity compared to the Fermi velocity,
we have an effective suppression
(for energy transfer rates) of $\sim \left(\frac{k}{p_F}\right)^4 \frac{m_n^3}{\mu^2 m_X}$,
where $m_X$ is the mass of the scatterer, $m_n$ is
the effective mass of the quasi-particle, and $\mu = m_X m_n / (m_X + m_n)$ is
the reduced mass.

Since the cross sections for scattering off single-particle excitations
also add incoherently, $\sigma \propto A$, coherent scattering should
dominate unless the size of the state is very large, so that as reviewed
in the previous sub-section, the coherent transfer cross section grows
slower than $A$. Making an estimate for SM-like nuclear matter
(individual constituents of mass $m_1 = 1 \GeV$, bulk nuclear
matter density $\rho = 1 \GeV / \fermi^3$) scattering off solar material,
we would need, very roughly, $A \gtrsim 10^{32}$ for incoherent scattering
to start dominating the transfer cross section (corresponding to
$R \gtrsim 40 \, {\rm \mu m}$). This is well above the size beyond
which thermalisations were found to be ineffective
in Section~\ref{sec:fusions}. However, it does illustrate why,
for example, scattering of DM in neutron stars is dominated by
incoherent scattering.


\section{Statistical identification of rising distributions}
\label{app:rising}

Suppose that we have some (one-dimensional) data points,
which we assume are IID samples from some probability distribution.
We wish to test the hypothesis that this distribution is
non-increasing, with respect to some appropriate function of the parameter.

If we expect plausible alternative distributions to feature only
one prominent rising segment (for example,
the recoil spectra considered in Section~\ref{sec:elastic} have successive Bessel function
peaks suppressed by both the natural fall-off of the Bessel function,
and the velocity distribution), a sensible approach is 
to locate the interval with the `worst' bias towards its right-hand
end, and ask what the probability is that a given sample
from a candidate non-increasing distribution would have 
an interval `that extreme'.\footnote{This would be a poor approach
if we expected the rises in the distribution to be e.g.\ a
small periodic signal super-imposed on some larger background, or more
generally any small but structured deviation from a larger background.
}

A simple measure of how biased an interval is towards its right-hand
end is simply the average position of the points within it. The
non-increasing distribution maximising the probability of right-biased
points is clearly the uniform distribution on the interval.
If observations are binned, we can find the exact distribution
for the average of bin mid-point positions, for some number $n$
of samples, by performing the convolution of the (binned) uniform
distribution on the interval with itself $n$ times. In particular, if the bins
are of uniform widths, this distribution can be computed analytically.
If observations are not binned, then the distribution for
the average of the positions is the (rescaled) Irwin-Hall distribution
for $n$ points. In both the discrete and continuous cases, the
null distribution is approximately normal for large $n$, with
variance $1/\sqrt{12 n}$ (on the interval $[0,1]$).

So, if we have $n$ samples with a mean position of $x$
(rescaling the interval to be of width 1), we can use the appropriate
distribution to find the probability of a mean position $\ge x$
arising from a uniform distribution on that interval. The
test statistic for the whole sample is then the minimum such
$p$-value for each sub-interval within the sample (since
it is clear that the worst sub-interval will always terminate
either at the left sample end, or at the location of a point, there
are a finite number of sub-intervals to test).
To determine the distribution of this test statistic under
a candidate non-increasing distribution, Monte Carlo
simulation can be used.
A uniform distribution is clearly the non-increasing distribution
most likely to produce fake rises, and while the data may indicate
that the underlying distribution is far from uniform, 
for reasonably small numbers of samples the $p$-values
obtained by adopting a uniform null distribution are only
slightly worse than those obtained by allowing
a free null distribution constrained by the fit to data.
In particular, we adopt a uniform null distribution for the calculations
in Figure~\ref{fig:fallingp}, and find that the $p$-values are generally
no more than a factor $\sim 2$ worse.

One issue is that, in many physical cases, there will be some
resolution associated with our points. Then, the distribution
our samples are drawn from is some (positive) underlying distribution convolved
by a (positive) detector response function, so must be smooth
on scales of order the resolution. This means that rightwards bunching
on such scales must be spurious, so should not be considered
in our test. The simplest way to solve this issue, if we expect
rising features in plausible alternative distributions
to be on scales larger than the resolution, is to bin the points
on around the resolution scale. 

Quantitatively, the variance for the average of $n$ uniform random
variables on $[0,1]$ is $\frac{1}{\sqrt{12 n}}$, and the expected
average position from a linearly rising distribution is $\frac{1}{2} +
\frac{1}{6}$, so we would expect to exclude the uniform distribution at
a significance of $\sim 0.6 \sqrt{n}$ sigma. In the example
from Figure~\ref{fig:fallingp}, with 100 events overall,
we expect around 26 in the first rise, so we would expect to obtain
around a $3 \sigma$ exclusion, assuming that the worst intervals arising
from the null distribution are of approximately that many events.
This is indeed what we observe, with
the $p$-value CDF in the 100-sample
case following very closely this normal approximation
(the yellow curve in Figure~\ref{fig:fallingp}, lying almost entirely
underneath the yellow points). The CDFs
for the 30 and 50 sample cases are not so well approximated,
since the `worst interval' is more variable with small numbers of events.


\section{Solar capture of heavy WIMPS}
\label{app:solarcapture}

Here, we give a very brief review of the capture
of weakly-interacting, massive dark matter particles 
by the Sun.
For $X$ particles much heavier than any of the relevant SM
nuclei within the Sun (here, this means $m_X \gtrsim 400 \GeV$),
the rate of capture by the Sun scales as $C_\sun \sim
\sigma_{Xn} \rho_X m_X^{-2}$, where $\sigma_{Xn}$ is the scattering cross section with
SM nucleons (assumed to be elastic
and spin-independent)~\cite{Gould1987,
Gould1991}.\footnote{
One factor of $m_X^{-1}$ comes from the 
number density $n_X = \rho_X m_X^{-1}$, while another arises
from the fact that heavier particles lose a smaller fraction 
of their kinetic energy in collisions with SM nuclei,
so only the low-speed part of the WIMP velocity distribution can be captured.
This also assumes that $\sigma_{Xn}$ is low enough that
the Sun is optically thin to $X$ particles, which is
the case for $\sigma_{Xn} \lesssim 10^{-3} \pb$.}
Summing over capture rates from the various elements
in the Sun (the main contributions are from the CNO elements)
using a standard solar model~\cite{Bahcall2004}, we obtain
\begin{equation}
C_\sun \simeq 3 \times 10^{19} \second^{-1} 
\left(\frac{\sigma_{Xn}}{10^{-8} \pb}\right)
\left(\frac{1 \TeV}{m_X}\right)^2 ,
\label{eq:csun}
\end{equation}
(taking the couplings to protons and neutrons
to be the same).
We have assumed that the DM velocity distribution follows
the Standard Halo Model---since, for high-mass WIMPs, 
only the low-velocity part of the distribution can be captured
by scattering events, modifications that affect the low-velocity
distribution will alter the capture rate.
direct detection experiments imply that $\sigma_{Xn} \lesssim
10^{-8} \pb \frac{m_X}{1 \TeV}$~\cite{LUX2013}, and the age of the Sun is
$t_\sun \simeq 5 \Gyr$, so the number of $X$ particles captured
is $\lesssim 5 \times 10^{36} \left(\frac{1 \TeV}{m_X}\right)$.
As a fraction of the total flux of $X$ particles hitting the Sun,
we capture
$\sim 3 \times 10^{-7} \left(\frac{\sigma_{Xn}/m_X}{10^{-8} \pb / \TeV}\right)$.

Here, `captured' means that the $X$ particles are in gravitationally bound
orbits passing through the Sun. Subsequent scatterings with material in the Sun
will reduce the size of these orbits further, and eventually the
$X$ particles will (ignoring other interactions) settle into an
isothermal distribution
$\rho_X \sim e^{-r/(2 r_\star^2)}$, with~\cite{Griest1986}
\begin{equation}
r_\star = \left(\frac{3 T_\sun}{4 \pi G m_X \rho_\sun}\right)^{1/2} \simeq
2 \times 10^{-3} R_\sun \left(\frac{\TeV}{m_X}\right)^{1/2}
\left(\frac{T_\sun}{10^7 \kelvin}\right)^{1/2}
\left(\frac{150 \gram \cm^{-3}}{\rho_\sun}\right)^{1/2} \,,
\end{equation}
(where the temperature and density are appropriate to the solar core).
If $m_X \ll 100 \TeV$, then the initial orbits
will be small enough that planetary perturbations can be mostly
neglected, and then thermalisation occurs in less than the lifetime
of the Sun if 
$\sigma_{Xn} \gtrsim 3 \times 10^{-13} \pb \left(m_X/\TeV\right)^{3/2}$~\cite{Peter2009}.
For larger $m_X$, most of the initial orbits are large
enough that the effect of Jupiter perturbs them so that they no longer
pass through the Sun, resulting in most of them never thermalising~\cite{Peter2009}.

Finally we remark that the above calculations apply to the case of capture by elastic
scattering with SM nuclei.  If there are sufficiently low-lying excitations
that inelastic scatterings are possible, in either
or both direct detection experiments and solar capture,
this may change the possible parameter space. (See
\cite{Nussinov2009,Menon2009} for related investigations of solar
capture of inelastic DM.)



\bibliography{ndm}
\bibliographystyle{JHEP}

\end{document}